# A Self-Consistent Field Study of Interfacial Dynamics in Unentangled Homopolymer Fluids in a Sheared Channel

*Maja Mihajlovic*

*Department of Chemistry, City College, City University of New York, New York, NY 10031, USA*

*Tak Shing Lo*

*The Levich Institute, City College, City University of New York, New York, NY 10031, USA*

*Yitzhak Shnidman\**

*Department of Engineering Science and Physics, College of Staten Island, City University of New York, Staten Island, NY 10314, USA*

November 10, 2004.

PASC: 83.80.Tc, 83.80.Sg, 83.85.Pt, 83.50.Lh.

*Corresponding author. Electronic address: shnidman@mail.csi.cuny.edu

# ABSTRACT

In a preceding paper, we have presented a general lattice formulation of the dynamic self-consistent field (DSCF) theory for inhomogeneous, unentangled homopolymer fluids. Here we apply the DSCF theory to study both transient and steady-state interfacial structure, flow and rheology in a sheared planar channel containing either a one-component melt or a phase-separated, two-component blend. We focus here on the case that the solid-liquid and the liquid-liquid interfaces are parallel to the walls of the channel, and assume that the system has translational symmetry within planes parallel to the walls. This symmetry allows us to derive a simplified, quasi-one-dimensional (quasi-1D) version of the DSCF evolution equations for free segment probabilities, momentum densities, and the ideal-chain conformation tensor. Numerical solutions of the quasi-1D DSCF equations are used to study both the transient evolution and the steady-state profiles of composition, density, velocity, chain deformation, stress, viscosity and normal stress within layers across the sheared channel. Good qualitative agreement is obtained with previously observed phenomena.

# I. INTRODUCTION

Processing of thermoset plastics typically involves mixing of solid pellets, composed of different polymers species, and having different sizes and shapes, within rotating screw extruders. As they are transported by the rotating screws, the pellets are melted, and intense stresses are generated by momentum transfer from the rotating screws. This results in a distribution of liquid domains of nearly homogeneous composition that are suspended in, but immiscible with, a majority liquid phase at a different homogeneous composition. These liquid domains are separated from each other, and from the surrounding solid walls, by narrow, generally curved, interfacial regions with steep composition gradients. The intense stresses induce nonuniform viscoelastic flows advecting and deforming the interfacial boundaries. Local rheology of such inhomogeneous polymer fluids is a function of chain stretching and orientation, in response to nonuniform flows and to thermodynamic forces that are generated by the compositional inhomogeniety at the interface. The material properties of the product, produced by melt extrusion through a shaping dye and subsequent solidification, exhibit strong dependence on interfacial composition, stresses and chain conformations induced during melt shaping.

Quantitative modeling of evolving interfacial structure, velocity and rheology in sheared, inhomogeneous polymer fluids is a challenging problem. Understanding how the interfacial properties are coupled dynamically to chain conformations and their deformation and orientation under flow is of great importance for polymer processing and for many other industrial and biological applications.

Realistic modeling of interfacial dynamics in polymer processing is exceedingly complex on several counts. Polymer fluids exhibit non-Newtonian rheology and viscoelastic flows arising from competition between chain deformation by nonuniform flows and entropically driven chain relaxation. The latter is constrained by chain entanglements when polymer chains exceed a characteristic entanglement length, which is the common situation in polymer processing. Much progress was achieved in understanding and modeling the rheology of *homogeneous* unentangled and entangled polymer fluids improving upon the Rouse and the reptation models, respectively [1-5]. The assumption of homogeneous composition allows neglect of species transport. However, species transport plays an important role in *inhomogeneous* polymer fluids where the evolution of velocity and stress fields across narrow interfacial regions is coupled to the evolution of the composition field. This raises the computational cost, since such coupling involves disparate length and time scales. Steep variations in composition and velocity fields across narrow interfacial regions require spatial resolution on the scale of the Kuhn length. On the other hand, shape deformation and motion of individual domains occurs on a much larger length scale. Similarly, the characteristic time for viscous momentum propagation in such systems is much shorter than the characteristic times for diffusive transport and for chain relaxation.

In general, the composition, velocity and stress fields during processing of immiscible polymer blends lack spatial symmetry, and thus require computations and analysis on three-dimensional grids. To reduce the computational cost barrier for validating quantitative models for polymer processing, it is advantageous to study simplified processes for melts and blends where the coupled evolution of composition, velocity and stress fields across narrow interfacial regions exhibits spatial symmetry.

Examples of such simplified process are provided by a one-component melts (Fig. 1(a)) or two component immiscible blends of homopolymers (Fig. 1(b)), confined in a channel between two planar walls. At an initial time, the melt or blend phases are equilibrated with solid-liquid interfaces adjacent and parallel to each wall. In the blend case, there is also a liquid-liquid interface separating the two coexisting equilibrium phases, which is parallel and centered midway between the two walls. The two walls are then sheared at constant, but opposite velocities, until nonequilibrium steady-state is reached. The objective is to study the coupled time evolution and the steady-state profiles of composition, density, velocity, chain deformation, stress, viscosity and normal stress within layers across the sheared channel.

These are relatively simple examples of interfacial dynamics in polymer processing, that have been studied both experimentally, and using a variety of models and simulation methods. Experimentally, it is known that at low shear rates such systems exhibit translational invariance with in planes parallel to the walls, though at high shear rates instabilities generated at the solid-liquid interface may break this symmetry [6]. Such symmetry allows using a quasi-one-dimensional (quasi-1D) spatial grid in a computational model for these processes at low shear rates, thus greatly reducing the computational cost.

Two-component immiscible homopolymer blends in a sheared planar channel exhibit a characteristic velocity slip and reduction of the shear viscosity at the liquid-liquid interface. This phenomenon has been first predicted based on scaling arguments by de Gennes and co-workers [7,8]. More recently, Goveas and Fredrickson [9] analyzed this phenomenon in symmetric, unentangled polymer blends, using numerical solutions of approximate constitutive equations for the evolution of composition and the deviatoric stress. Subsequently, Barsky and Robbins published a very detailed molecular dynamics study of interfacial slip in symmetric, unentangled polymer blends [10] and related it to the chain conformation at the sheared interface [11]. Most recently, similar velocity slip phenomena have been observed experimentally [12-14], though the bulk phases of the observed polymer fluids were in the entangled regime.

In a preceding paper [15], we presented a general, three-dimensional formulation of a *dynamic* self-consistent field (DSCF) lattice theory that is capable of modeling interfacial dynamics in inhomogeneous, compressible polymer fluids composed of *unentangled* homopolymer chains in a sheared channel. Self-consistent field (SCF) theories *approximate* the configurational probability for a many-body system with a *product of one-body probabilities* for each system constituent, interacting solely with a conjugate mean external field. For simple fluids, the constituents are atoms or molecules, and the mean external field represents interactions with other atoms or molecules and walls, *averaged over their one-body probability distributions.* Polymer fluids include macromolecules that are approximated as freely jointed chains of Kuhn segments. Thus Kuhn segments, rather than entire macromolecules are the basic constituents in polymer SCF theory. The latter approximates the joint configurational probability for all the Kuhn segments by a product of one-body probabilities for "free" (unconnected) segments in a self-consistent mean field representing interactions with adjacent segments. These interactions are averaged not over the one-body free segment probabilities at the adjacent sites, but rather over local segmental volume fractions at those adjacent sites, which account for intrachain correlations between segments belonging to the same chain. In a homogeneous polymer melt at equilibrium, the chains are considered ideal (noninteracting), and are modeled by isotropic random walks on a lattice. In *inhomogeneous* polymer melts and blends at equilibrium, chain conformations are modeled as anisotropic lattice random walks *in a self-consistent field* [16].

Mean-field theories for *simple* fluids *at equilibrium* determine the composition of interfacial regions by balancing the molecular interaction contribution to the free energy (modeled by a self-consistent field potential acting on each molecule) with the entropy of mixing. *Minimization of the free energy* determines the steep variation of the compositional profile across the interface. Its width typically extends over only a few molecular diameters. Equilibrium polymer SCF theories account, in addition, for the entropic contributions to the free energy arising from intrachain correlations, and for the enhanced probability for the occurrence of certain chain conformations which are favored by the self-consistent field representing the interactions with adjacent segments and walls [16].

As in equilibrium SCF approaches, our DSCF theory approximates the configurational probability for all segments in the system by the product of one-body free segment probabilities interacting with a self-consistent mean field representing *connected* segments and walls. However, in DSCF both the one-body free segment probabilities and the local self-consistent fields are functions of time. DSCF models the effect of nonuniform flow on ideal chain conformations with the FENE-P dumbbell model for the time evolution of the conformation tensor (the second moment of the chain end-to-end distance), and relates it to the (conditional) stepping probabilities in the lattice random walks representing the chain conformations. Free segment and stepping probabilities generate statistical weights for chain conformations in a self-consistent mean potential field, and determine the local volume fractions of *connected* segments. Flux balance across a unit lattice cell yields mean-field transport equations for free segment probabilities and momentum densities, determining their time evolution. Diffusive and viscous contributions to the fluxes arise from segmental hops modeled as a Markov process. Hopping transition rates depend on the changes in the free energy, reflecting segmental interactions, kinetic energy, and entropic contributions accounting for chain deformation under flow. The DSCF theory results in a system of ordinary differential equations (ODEs) evolving a set of independent variables defined at the sites of a face-centered cubic (FCC) lattice. The independent variables include the free segment probabilities for each segmental species, the mass-averaged momentum density, and the ideal-chain conformation tensors for each of the chain species. Other physical quantities needed for analyzing the evolving interfacial structure, velocity field, rheology, and the statistics of chain conformations can be expressed as functions of these independent variables. A detailed three-dimensional formulation of the DSCF theory, complete with physical reasoning for each of its components and a discussion of their relation to prior work, can be found in the preceding paper [15].

In this paper we present the first application of our DSCF theory to study interfacial structure, velocity and rheology in one-component melts and in phase-separated, two-component blends, sheared between two planar walls. To reduce the computational cost, we focus on polymer fluids exhibiting translational invariance throughout the shearing process. The remainder of the present paper is structured as follows. In Section II we present a simplified, quasi-one-dimensional (quasi-1D) version of the DSCF evolution equations for free segment probabilities, momentum densities, and the ideal-chain conformation tensors, obtained from the general DSCF theory formulated in the preceding paper [15] under the simplifying assumption of translational invariance within layers parallel to the shear walls. In the next two sections we present numerical solutions of the quasi-1D DSCF equations, and use these solutions to analyze both the transient evolution and the steady-state profiles of the segmental volume fractions, velocity field, shear stress, chain deformation by flow, and first normal stress difference across the channel. In Section III we use the quasi-1D DSCF solutions to study a one-component melt of unentangled homopolymer chains in a channel between two sheared planar walls. In Section IV we use the quasi-1D DSCF equations to study a phase-separated, two-component blend of unentangled homopolymer chains in a sheared planar channel. Good qualitative agreement is obtained with previously observed phenomena, such as chain depletion next to the walls and interfaces, and the associated velocity slip and reduced viscosity in interfacial regions. Conclusions and perspectives for further application, validation and extension of our new DSCF theory are presented in Section V.

## II. DSCF EVOLUTION EQUATIONS WITH PLANAR SYMMETRY

In this paper we apply the DSCF theory to unentangled polymer fluids in a channel between two sheared planar walls (Figs. 1(a) and 1(b)). As in the preceding paper [15], polymer chains are modeled as freely jointed Kuhn segments of length $a$, and it is assumed that the Kuhn segments are confined to Wigner-Seitz cells of an FCC lattice (Fig. 1(c)) with its lattice constant being the Kuhn length. However, here we focus on systems that are invariant under translations in the $xy$-plane, such as one component melts (Fig. 1(a)) and two-component phase-separated blends (Fig. 1(b)), with planar interfaces parallel to the walls. This means that all quantities at sites $\mathbf{r}$ belonging to the same triangular lattice layer $i$ parallel to the walls are identical (Fig. 1). The spatial computational grid for such a layered system is one-dimensional, and we now proceed to simplify considerably the DSCF equations presented in the preceding paper [15], by taking advantage of this translational symmetry.

Let $P_i^\alpha$ be the one-body probability for a free (disconnected) Kuhn segment of type $\alpha$ to occupy a site in layer $i$ ($i = 1, 2, \ldots, L$, where $L$ is the number of lattice layers between the two walls in Fig. 1). For a one-component melt consisting of homopolymers of type $A$, $\alpha = A$, while for a blend consisting of two homopolymer components $A$ and $B$, $\alpha = A$ or $B$. Conformations of interacting chains are modeled by lattice random walks in a self-consistent field. The segmental volume fraction of type-$\alpha$ segments at a site in layer $i$ that belongs to a chain of $N^\alpha$ freely jointed Kuhn segments, interacting with walls and other chains, is [16]

$$\phi_i^\alpha = \frac{\overline{\phi}^\alpha L}{N^\alpha \sum_{\mathbf{r}} P_{\mathbf{r}}^\alpha \left( N^\alpha \right)} \sum_{s=1}^{N^\alpha} \frac{P_i^\alpha \left( s \right) P_i^\alpha \left( N^\alpha - s + 1 \right)}{P_i^\alpha \left( 1 \right)} \tag{1}$$

where $P_i^\alpha \left( s \right)$ are statistical weights for finding a terminus of an $\alpha$-type chain of $s$ freely jointed Kuhn segments, given by the one-dimensional master equation

$$P_i^\alpha \left( s \right) = P_i^\alpha \left( 1 \right) \left[ \lambda_{+,i-1}^\alpha P_{i-1}^\alpha \left( s \right) + \lambda_{0,i}^\alpha P_i^\alpha \left( s \right) + \lambda_{-,i+1}^\alpha P_{i+1}^\alpha \left( s \right) \right], \tag{2}$$

for $s = 2, 3, \ldots, N^\alpha$ and $P_i^\alpha (1) = P_i^\alpha / \left( 1 - P_i^A - P_i^A \right)$ is the statistical weight for a free Kuhn segment (monomer). In Eq. (1), $\bar{\phi}^\alpha$ is the average volume fraction of species $\alpha$ in the system and in Eq. (2), $\lambda_{\pm,i}^\alpha$ is the stepping probability from a site in layer $i$ to an adjacent site in layer $i \pm 1$ in a lattice random walk representing the conformation of an ideal chain of type $\alpha$, whereas $\lambda_{0,i}^\alpha$ is the stepping probability from a site in layer $i$ to an adjacent site in the same layer. In the notation of Ref. [15],

$$\lambda_{0,i}^\alpha = 2 \left( \lambda_{\mathbf{a}_1,\mathbf{r}}^\alpha + \lambda_{\mathbf{a}_2,\mathbf{r}}^\alpha + \lambda_{\mathbf{a}_3,\mathbf{r}}^\alpha \right) \quad \text{and} \quad \lambda_{+,i}^\alpha = \lambda_{\mathbf{a}_7,\mathbf{r}}^\alpha + \lambda_{\mathbf{a}_8,\mathbf{r}}^\alpha + \lambda_{\mathbf{a}_9,\mathbf{r}}^\alpha = \lambda_{-,i}^\alpha, \quad (3)$$

where the site $\mathbf{r}$ belongs to layer $i$, and $\lambda_{\mathbf{a}_k,\mathbf{r}}^\alpha$ is the conditional probability for placing the next segment at one of the sites site $\mathbf{r} + \mathbf{a}_k$ ($k = 1, 2, \ldots, 12$), that are adjacent to $\mathbf{r}$ on the FCC lattice in the lattice random walk generating a chain conformation. As in [15], we assume a reflection symmetry: $\lambda_{\mathbf{a}_l,\mathbf{r}}^\alpha = \lambda_{\mathbf{a}_k,\mathbf{r}}^\alpha$ if $\mathbf{a}_l = -\mathbf{a}_k$. The three layer stepping probabilities are related linearly to the components of conformation tensor $\mathbf{S}_i^\alpha$ for ideal chains under flow:

$$\lambda_{0,i}^\alpha = \frac{1}{(N^\alpha - 1) a^2} \left( S_{xx,i}^\alpha + S_{yy,i}^\alpha - \tfrac{1}{2} S_{zz,i}^\alpha \right) \quad \text{and} \quad \lambda_{-,i}^\alpha = \lambda_{+,i}^\alpha = \frac{1}{(N^\alpha - 1) a^2} \tfrac{3}{4} S_{zz,i}^\alpha . \quad (4)$$

where $\mathbf{S}_i^\alpha$ evolves in time according to the FENE-P dumbbell model,

$$\overset{\triangledown}{\mathbf{S}}_i^\alpha = -\frac{1}{\tau_{db,i}^\alpha} \left[ \frac{\mathbf{S}_i^\alpha}{1 - \frac{3}{(N^\alpha - 1) \tilde{a}^2 b^\alpha} \mathrm{Tr} \mathbf{S}_i^\alpha} - \frac{\left( N^\alpha - 1 \right) \tilde{a}^2}{3} \boldsymbol{\delta} \right], \quad (5)$$

and $\overset{\triangledown}{\mathbf{S}}_i^\alpha$ denotes the upper-convected time derivative of the tensor $\mathbf{S}_i^\alpha$,

$$\overset{\triangledown}{\mathbf{S}}_i^\alpha \equiv \frac{\partial \mathbf{S}_i^\alpha}{\partial t} + \mathbf{u}_i \cdot \nabla \mathbf{S}_i^\alpha - \nabla \mathbf{u}_i^{\mathsf{T}} \cdot \mathbf{S}_i^\alpha - \mathbf{S}_i^\alpha \cdot \nabla \mathbf{u}_i . \quad (6)$$

Here $\tau_{db,i}^{\alpha} = N^{\alpha}\left(N^{\alpha}-1\right)\tilde{a}_{\alpha}^{2}\zeta_{i}^{\alpha}/24k_{B}T$ is the local relaxation time for a Hookean dumbbell with the spring constant $3k_{B}T/\left[\left(N^{\alpha}-1\right)\tilde{a}_{\alpha}^{2}\right]$, modeling ideal chains of type $\alpha$, and $\tilde{a}_{\alpha}=c_{\alpha}a$, where $c_{\alpha}=\sqrt{\left(b^{\alpha}+3\right)/b^{\alpha}}$. The latter relations fine-tunes the spring

constant used in the FENE-P model, so that the diagonal components of $\mathbf{S}_{i}^{\alpha}$ obtained

from the equilibrium solution of Eq. (6) revert to the known equilibrium value

$\left(N^{\alpha}-1\right)a^{2}/3$ for the diagonal components of $\mathbf{S}_{i}^{\alpha}$ for a chain consisting of $N^{\alpha}$ Kuhn

segments, joined together by $N^{\alpha}-1$ freely jointed links, of length $a$ each. This reproduces isotropic ,equilibrium stepping probabilities $\lambda_{\mathbf{a}_{k},i}^{\alpha}=\frac{1}{12}$ for stepping along any

of the twelve unit vectors $\mathbf{a}_{k}$ connecting between a site in layer $i$ and its nearest

neighbors on the FCC lattice, as is assumed in the equilibrium SCF theory of Scheutjens

and Fleer [17] and its compressible variants [18-20]. In this case Eqs. (1) and (2) are

identical to those provided by a compressible variant [20] of the Scheutjens-Fleer SCF

theory. For isothermal polymer fluids subjected to stresses, the free segment probabilities

$P_{i}^{\alpha}$ and the local momentum densities $\mathbf{g}_{i}^{\alpha}$ of species $\alpha$ evolve according to the one

dimensional transport equations

$$\frac{dP_{i}^{\alpha}}{dt}=-\nabla\cdot\left(P_{i}^{\alpha}\mathbf{u}_{i}\right)-\left(\frac{\mathbf{j}_{i}^{\alpha}-\mathbf{j}_{i-1}^{\alpha}}{\sqrt{2/3}a}\right)\cdot\mathbf{e}_{3}, \tag{7}$$

and

$$\frac{d\mathbf{g}_{i}^{\alpha}}{dt}=-\nabla\cdot\left(\mathbf{g}_{i}^{\alpha}\mathbf{u}_{i}+\boldsymbol{\varepsilon}_{i}^{\alpha}\right)-\left(\frac{\boldsymbol{\pi}_{i}^{\alpha}-\boldsymbol{\pi}_{i-1}^{\alpha}}{\sqrt{2/3}a}\right)\cdot\mathbf{e}_{3}-\frac{\phi_{i}^{\alpha}}{wP_{i}^{\alpha}}\left(\zeta_{i,i+1}^{\alpha}\mathbf{j}_{i}^{\alpha}+\zeta_{i,i-1}^{\alpha}\mathbf{j}_{i-1}^{\alpha}\right). \tag{8}$$

Eqs. (7) and (8) are the one-dimensional versions of the general DSCF transport Eqs. (28) and (57) in [15], simplified using the translational symmetry within $xy$-layers. They can be also obtained independently, by balancing free segment probability and momentum density fluxes over a control volume of $w = a^3 / \sqrt{2}$ consisting of a rectangular lattice unit cell. The coefficient $\zeta_{i,i+1}^\alpha = \sqrt{\zeta_i^\alpha \zeta_{i+1}^\alpha}$ is the geometric average of the local segmental friction coefficients in layers $i$ and $i+1$ obeying Doolitles' law [15,21], and $\mathbf{u}_i = w \sum_\alpha \mathbf{g}_i^\alpha / \sum_\alpha m^\alpha \phi_i^\alpha$ (where $m^\alpha$ is the segmental mass of species $\alpha$) is the mass-averaged velocity at a site in layer $i$. The diffusive free segment probability current and the viscous stress are given by

$$
\begin{aligned}
\mathbf{j}_i^\alpha = 3\sqrt{\tfrac{2}{3}} \left(1 - \delta_{i+1,1}\right)\left(1 - \delta_{i,L}\right) \frac{D_{i,i+1}^\alpha}{\left(1 - \bar{\phi}\right)} \\
\times \left[ P_i^\alpha \left(1 - \phi_{i+1}\right) \varphi\!\left(\frac{\Delta\left\langle \mathcal{H}_i^\alpha \right\rangle}{k_B T}\right) - P_{i+1}^\alpha \left(1 - \phi_i\right) \varphi\!\left(-\frac{\Delta\left\langle \mathcal{H}_i^\alpha \right\rangle}{k_B T}\right) \right] \frac{\mathbf{e}_3}{a}
\end{aligned}
\tag{9}
$$

and

$$
\begin{aligned}
\boldsymbol{\pi}_i^\alpha = 3\sqrt{\tfrac{2}{3}} \left(1 - \delta_{i+1,1}\right)\left(1 - \delta_{i,L}\right) \frac{\nu_{i,i+1}^\alpha}{\left(1 - \bar{\phi}\right)} \\
\times \left[ \mathbf{g}_i^\alpha \left(1 - \phi_{i+1}^\alpha\right) \varphi\!\left(\frac{\Delta\left\langle \mathcal{H}_i^\alpha \right\rangle}{k_B T}\right) - \mathbf{g}_{i+i}^\alpha \left(1 - \phi_i^\alpha\right) \varphi\!\left(-\frac{\Delta\left\langle \mathcal{H}_i^\alpha \right\rangle}{k_B T}\right) \right] \frac{\mathbf{e}_3}{a}.
\end{aligned}
\tag{10}
$$

Here the locally averaged self-diffusivity and kinematic viscosity of species $\alpha$ are given by

$$
D_{i,i+1}^\alpha = \frac{k_B T}{N^\alpha \zeta_{i,i+1}^\alpha} \qquad \text{and} \qquad \nu_{i,i+1}^\alpha = \frac{N^\alpha \zeta_{i,i+1}^\alpha a^2 b^\alpha}{24 m^\alpha \left(b^\alpha + 3\right)},
\tag{11}
$$

respectively, where $b^\alpha = 3\left(N^\alpha - 1\right)$ is the finite extensibility parameter in the FENE-P dumbbell model for polymer chains. In Eqs. (3) and (4), the function $\varphi$ is a transition rate function satisfying local detailed balance, and $\Delta\left\langle\mathcal{H}_i^\alpha\right\rangle = \left\langle\mathcal{H}_{i+1}^\alpha\right\rangle - \left\langle\mathcal{H}_i^\alpha\right\rangle$ is the change in the free energy caused by a segment of species $\alpha$ hopping from a site in layer $i$ to an adjacent vacant site in layer $i+1$. The local contribution to the free energy of a segment of species $\alpha$ in layer $i$ is

$$\left\langle\mathcal{H}_i^\alpha\right\rangle = k_B T\left[\frac{1}{2}\sum_{\beta=A,B}\left(1-\delta_{\alpha\beta}\right)\chi_{AB}\left\langle\left\langle\phi_i^\beta\right\rangle\right\rangle - \chi_s^\alpha\left(\delta_{i1} + \delta_{iL}\right)\right] + \frac{1}{2}m^\alpha\phi_i^\alpha\mathbf{u}_i^2$$
$$- \frac{\phi_i^\alpha k_B T}{2N^\alpha}\left\{\mathrm{tr}\left(\boldsymbol{\delta} - \frac{3}{\left(N^\alpha-1\right)\tilde{a}_\alpha^2}\mathbf{S}_i^\alpha\right) + \ln\left[\det\left(\frac{3}{\left(N^\alpha-1\right)\tilde{a}_\alpha^2}\mathbf{S}_i^\alpha\right)\right]\right\}.$$

(12)

The first term on RHS of the equation above accounts for interactions between a pair of nearest-neighbor segments of distinct homopolymer components, and vanishes in a one-component homopolymer melt. It is characterized by the segment-segment interaction parameter $\chi_{\alpha\beta}$, that is closely related to the Flory-Huggins interaction parameter $12\chi_{\alpha\beta}$, the latter defined as the energy change (in units of $k_B T$) due to the transfer of an $\alpha$ segment from a solution of pure $\alpha$ to a solution of pure $\beta$. For segments of the same size, it is assumed that $\chi_{AA} = \chi_{BB} = 0$ and that $\chi_{AB} = \chi_{BA}$. The double angular brackets represent summation over all nearest neighbors of a site in layer $i$

$$\left\langle\left\langle\phi_i^\beta\right\rangle\right\rangle = 3\left(1-\delta_{i1}\right)\phi_{i-1}^\beta + 6\phi_i^\beta + 3\left(1-\delta_{iL}\right)\phi_{i+1}^\beta.$$

(13)

The second term represents interactions between the segments in the first (or last) layer and the solid wall. The wall interaction parameter $\chi_s^\alpha$ is defined as the energy change (in units of $k_B T$) due to the transfer in a pure-$\alpha$ fluid of an $\alpha$ segment to a layer adjacent to a wall from a layer further away from the wall. It is a positive for attractive segment-wall interactions, and negative for repulsive segment-wall interactions.

The third term is the kinetic energy contribution arising from the local mean convective velocity. The last term depends on the second moment $\mathbf{S}_i^\alpha = \left\langle \mathbf{Q}_i^\alpha \mathbf{Q}_i^\alpha \right\rangle$ of the end-to-end distance $\mathbf{Q}_i^\alpha$ of an ideal (noninteracting) FENE-P chain deformed by a nonuniform flow in a homogeneous fluid, with its center of mass located in layer $i$. It accounts for the contribution to the local free energy due to chain deformations caused by nonuniform flows [22].

In Eq. (8), $-\boldsymbol{\varepsilon}_i^\alpha$ is the elastic contribution to the stress tensor,

$$\boldsymbol{\varepsilon}_i^\alpha = \frac{\phi_i^\alpha \zeta_i^\alpha}{8} \left[ \overset{\triangledown}{\mathbf{T}}_i^\alpha + \frac{\left(N^\alpha - 1\right)\tilde{a}_\alpha^2}{3} \left( \frac{\partial M_i^\alpha}{\partial t}\boldsymbol{\delta} + \left(\mathbf{u}_i \cdot \nabla M_i^\alpha\right)\boldsymbol{\delta} \right) \right], \qquad (14)$$

where

$$M_i^\alpha = 1 - \frac{3}{\left(N^\alpha - 1\right)\tilde{a}_o^2 b^\alpha}\mathrm{Tr}\mathbf{S}_i^\alpha , \qquad (15)$$

and

$$\mathbf{T}_i^\alpha = \mathbf{S}_i^\alpha - \frac{\left(N^\alpha - 1\right)\tilde{a}_\alpha^2}{3}M_i^\alpha\boldsymbol{\delta} \qquad (16)$$

is the deviatoric chain conformation tensor. Note that Eq. (14) has transient terms that vanish at a steady state.

The evolution equation for the total momentum density is obtained by summing over the contributions of the two species,

$$\frac{d\mathbf{g}_i}{dt} = \sum_\alpha \frac{d\mathbf{g}_i^\alpha}{dt} . \qquad (17)$$

In order to solve the quasi-1D DSCF equations above, one has to prescribe boundary conditions for $\mathbf{g}_i$ (or equivalently, for $\mathbf{u}_i = \mathbf{g}_i / \rho_i$) at sites adjoining the walls. A connected segment of type $\alpha$ occupies a site in a layer $i$ adjacent to a wall ($i=1$ or $i=L$) with probability $\phi_i^\alpha$. If this happens, we assume that the connected segment acquires the velocity of the adjacent wall. Therefore, for the geometry specified in Figs. 1 and 2, the mass-averaged mean velocity at the sites next to the walls is set to

$$\mathbf{u}_1 = -\sum_\alpha \phi_1^\alpha u_w \mathbf{e}_1, \qquad \mathbf{u}_L = \sum_\alpha \phi_L^\alpha u_w \mathbf{e}_1. \tag{18}$$

Eqs. (5), (7) and (8) constitute a closed system of nonlinear ordinary differential equations of the form

$$\dot{\mathbf{X}} = \mathbf{F}\left[\mathbf{X}, \mathbf{Y}(\mathbf{X})\right] \tag{19}$$

where $\mathbf{X} = \left\{ P_i^\alpha, \mathbf{g}_i^\alpha, \mathbf{S}_i^\alpha \right\}_{i=1,\ldots,L}^\alpha$, and the auxiliary function $\mathbf{Y}(\mathbf{X})$ is defined by Eqs. (1)-(4) and (9)-(16). All other physical quantities of interest in our DSCF simulations are expressed as functions of the independent variables $\mathbf{X}$. We assume that polymer fluid is confined between two parallel solid walls, normal to the $z$-axis in a Cartesian system of coordinates, defined by a triad of unit vectors $\mathbf{e}_1 = (1,0,0)$, $\mathbf{e}_2 = (0,1,0)$ and $\mathbf{e}_3 = (0,0,1)$, as shown in Fig. 1. The two walls are moving at opposite, but constant velocities $\pm u_w \mathbf{e}_1$. The positive sign corresponds to the wall at $z = (L+1)\sqrt{2/3}a$, that is adjacent to the top layer $i=L$ at $z_L = L\sqrt{2/3}a$. The negative sign corresponds to the wall at $z=0$, that is adjacent to the bottom layer $i=1$ at $z_1 = \sqrt{2/3}a$. This results in a nominal shear rate for the two walls of $\dot{\gamma} = 2u_w \big/ \left[ (L+1)\sqrt{2/3}a \right]$.

# III. ONE-COMPONENT MELTS

## A. Simulation Set Up and Parameter Estimation

As the first application of the DSCF theory, we have studied the morphology and rheology of unentangled homopolymer melts consisting of a single component A in a sheared planar channel between two solid walls (see Fig. 1(a)). The DSCF model captures such steady-state effects as the depletion of polymer chains, and the velocity slip, in the vicinity of the solid walls. It also allows the examination of the time evolution of velocity, stress and chain conformation tensors, as momentum propagates from the walls at the onset of shear, and after the shear is stopped.

We assume that the homopolymer chains in the melt are linear and unentangled. Furthermore, we assume that the melt stays invariant under translations in directions parallel to the walls, and is kept isothermal and isobaric at temperature $T = 509 \, ^\circ \text{K}$ and atmospheric pressure. Each homopolymer chain is modeled by $N^\alpha$ freely jointed Kuhn segments. For the Kuhn segment length, we adopted the value $a = 4.6 \, \overset{\circ}{\text{A}}$ used by Li and Ruckenstein [23] in their equilibrium SCF study of polyethylene chains. The lattice constant of the FCC lattice used in our DSCF model is assigned the value of the Kuhn length cited above. The molecular weight of a homopolymer melt is assumed to be below the entanglement molecular weight, $M < M_e$. This restricts the number of Kuhn segments in a homopolymer chain to $N^\alpha < N_e$, where $N_e$ is the number of Kuhn segments at the between entanglements. The value of $N_e$ has been controversial. Estimates in the literature [24] vary between 35 and 75. Since the present form of our DSCF model is limited to unentangled fluids, we conservatively limited the number of Kuhn segments in the homopolymer chains to $N^\alpha \leq 32$, which is below the lower bound for $N_e$ cited above.

We used the Stokes-Einstein relation to determine the segmental friction coefficient from the value of the self-diffusion coefficient for an equilibrium polyethylene melt at the same temperature and pressure, obtained by Paul *et al* [25] from molecular dynamics (MD) simulations and neutron spin-echo spectroscopy. The free volume in the bulk of the melt is estimated from the difference between the densities of polyethylene at $T = 509K$ and at the glass transition temperature, $T_g = 263K$, both at atmospheric pressure. This leads to the estimate of the free volume fraction in the bulk phase of polyethylene melt at $T = 509K$ and atmospheric pressure being $\bar{\phi}^0 = 1 - \bar{\phi} = 0.150$, where $\bar{\phi} = 0.850$ is the volume fraction of polymer chains.

The quasi-1D DSCF theory is first used to equilibrate a homopolymer melt between static planar walls by solving the evolution equations with $\mathbf{g}_i^A = \mathbf{u}_i = 0$ at each layer, using the equilibrium layer stepping probabilities $\lambda_{0,i}^A = \frac{1}{2}$ and $\lambda_{\pm,i}^A = \frac{1}{4}$. The equilibration run was always started from the homogeneous bulk phase values for polyethylene melt at $T = 509K$ and atmospheric pressure. The results of the equilibration run provided the initial values for solving the quasi-1D DSCF evolution equations in a channel between two planar walls that are sheared at a constant nominal shear rate $\dot{\gamma}$ until we reached a nonequilibrium steady state. The procedure was iterated for a number of different values of $\dot{\gamma}$ and $N^\alpha$. To investigate relaxation of the nonequilibrium steady state back to thermodynamic equilibrium, in some instances we furthermore used the steady-state results as initial values for solving the quasi-1D DSCF equations at $\dot{\gamma} = 0$. The DSCF results reported below use a system of fundamental units consisting of $[energy] = k_B T$, $[length] = a$ and $[time] = \tau$, where $\tau = a^2 / D_0^{\alpha}$ is the characteristic for a probe (unconnected) segment of type $\alpha$ to diffuse a distance $a$, and $D_0^{\alpha}$ is the self-diffusion coefficient of the probe segment, related to its friction coefficient $\zeta_0^{\alpha}$ by the Stokes-Einstein relation $D_0^{\alpha} = k_B T / \zeta_0^{\alpha}$. All the quantities obtained from the DSCF simulations reported below, are expressed in these units.

**B. DSCF Simulation Results for Melts**

Fig. 2 shows the steady state segmental volume fraction profiles of the homopolymer melt across the channel, obtained at a nominal shear rate $\dot{\gamma} = 1 \times 10^{-3} \tau^{-1}$. The channel width is $L = 64\sqrt{2/3}a$. The linear polymer chains consist of $N^A = 24$ Kuhn segments. It is observed that, at this shear rate, the steady state segmental volume fractions do not differ noticeably from their equilibrium values. Here we vary the short-range interactions between the polymer segments and the solid walls between attractive ($\chi_s > 0$), neutral ($\chi_s = 0$) and repulsive ($\chi_s < 0$). All three profiles reveal the depletion of polymer segmental volume fraction in the layers next to the solid walls, relative to the bulk density. The degree of depletion depends on the imposed segment-wall interactions, as can be seen in Fig. 2. This depletion is followed by a non-monotonic, oscillatory approach towards bulk values, spanning the next few layers. If the system were infinite, the bulk values would correspond to the mean total volume fraction of the polymer fluid, $\bar{\phi} = 8.50 \times 10^{-1}$. However the bulk densities deviate from $\bar{\phi}$, due to the finite size effects caused by the small separation between the wall and the segment-surface interactions.

Note that if the original equilibrium SCF lattice theory of Scheutjens and Fleer [16] is applied to a one component melt at equilibrium between two static walls, the segmental volume fraction profile across the channel would be homogeneous, since their theory is incompressible. However, similar wall-depletion and finite-size effects are produced by compressible variants of Scheutjens and Fleer's equilibrium SCF theory [18-20]. It is also worth noting that a depleted layer adjacent to the wall forms as well in lattice gas models of simple liquids consisting of small molecules [26,27], if there is a strong repulsive interaction between the wall and the molecules in the adjacent layer. However, wall depletion of segmental volume fractions in polymer melts is predominantly caused by an entropic, rather than enthalpic effect, as the number of available chain conformations is reduced next to the walls. As shown in Fig. 2, wall depletion of segmental volume fractions in polymer melts occurs even when the walls are neutral ($\chi_s = 0$), though it can be enhanced by unfavorable wall interactions ($\chi_s < 0$) and diminished by favorable interactions with the wall ($\chi_s > 0$). The oscillatory decay of the depletion at increasing distances from the wall can be attributed to intrachain correlations between the segments.

Fig. 3 displays the time evolution of the $x$-component of the velocity profile, across the channel, obtained at a nominal shear rate $\dot{\gamma} = 1 \times 10^{-5} \tau^{-1}$. Note that we use a logarithmic time scale. The chain length is $N^{\alpha} = 24$, and the walls are neutral ($\chi_s = 0$). The initial state of the system (at time $t = 0$) corresponds to the system being at thermal equilibrium. Then, we set the velocities of the upper and lower walls to $\mathbf{u_w} = \pm u_w \mathbf{e}_1$, respectively. At subsequent times, we observe first rapid, non-monotonic propagation of the momentum (and, therefore, of the velocity) from the wall, which seems to be oscillatory in time (Fig.

4). This quasi-oscillatory behavior originate from $-\varepsilon_i^\alpha$, the elastic contribution to the stress (see Eqs. (14) - (16)), and it disappears when the elastic contribution to the stress is neglected in the momentum evolution equations. In a continuum PDE model of a sheared incompressible non-Newtonian fluid [28], a similar elastic contribution to the stress introduces a term of hyperbolic nature into the momentum evolution, giving rise to shock waves which are smeared and dampened by the parabolic viscous term. It is known that smearing of shock wave solutions to hyperbolic partial differential equations is enhanced by finite difference approximations of the type that we used here [29]. Hence it is possible that what seems as dampened oscillations in Figs. 3 and 4 should be interpreted as propagation of smeared shocked waves, which are reflected back and forth between the opposite walls.

At a longer time scale, these smeared shock waves or oscillations are damped out by the parabolic viscous term in the momentum transport equation (Eq. (8)), giving rise to a Couette-like steady state velocity profile. This profile is linear across the channel, except near the walls, where we observe a noticeable velocity slip. We note that velocity slip can be observed even in DSCF simulations of simple liquids of small molecules in a sheared capillary if there is a strong unfavorable interaction between the molecules and the walls that leads to a depletion layer at the walls [26,27]. Such velocity slip effects have been known since Tolstoi's pioneering observations [30] on capillary flow of water in hydrophobic micro-capillaries, and have been discussed extensively in the literature [31-33]. We believe that, in the case of polymer melts, the velocity slip at the wall (Figs. 3 and 4) is also caused by the depletion of the segmental volume fraction near the walls, as seen in Fig. 2, though in this case the depletion is dominated by the chain entropy effects mentioned above.

In Fig. 4 we focus on the lattice layer located at the lower quarter of the channel and show the development of velocity at this layer. One can notice that, at small times, the velocity at the layer is equal to 0. At $t \approx 10^{-1}\tau$ to $t \approx 3 \times 10^{-1}\tau$ it exhibits a slight undershoot, which may not be physical, but is rather an artifact of our finite-difference approximation to a partial differential equation of hyperbolic nature [29]. After this time, the smeared velocity shocked wave is reflected back and forth between the two channel walls, with the amplitude dampened by the viscous parabolic term. Finally, around $t \approx 2 \times 10^{2}\tau$ the shock waves or oscillations are completely dampened out, and the velocity acquires its steady state profile across the channel. Fig. 5 depicts the wave-like propagation of the velocity and its overshoot, across the channel, as it approaches the steady state. Mochimaru [28] studied the effect of elasticity on the unsteady-state development of plane Couette flow, using a constitutive equation for the FENE-P dumbbells. His results suggest that both the shear stress and velocity overshoot. This overshooting becomes more pronounced with increasing transient elasticity of the model fluid.

The time evolution of the shear stress, in a layer at the lower quarter-channel, is shown in Fig. 6. The chain length is $N^{A} = 24$ and the walls are assumed to be neutral ( $\chi_{s} = 0$ ). The system is first equilibrated and then subjected to a steady shear flow (at a nominal shear rate $\dot{\gamma} = 1 \times 10^{-5}\tau^{-1}$ ). Fig. 6(a) shows the development of the shear stress, at the layer located at the lower quarter of the channel, after the onset of shear. Until around $t \approx 7 \times 10^{-1}\tau$ the velocity has not propagated to this layer yet, and the shear stress is equal to 0. After this time, a slight undershoot develops in the shear stress before it starts rising. However, as explained above, this behavior could be an artifact arising from the well-known limitations of finite-difference approximation schemes for partial differential equation of hyperbolic nature [29], and if so, it should be disregarded. From this time on, the shear stress develops in the manner a smeared shock-wave with decaying amplitude. This is related to the shock-wave-like velocity propagation (Figs. 3 – 5). In the course of time, these shock waves are dampened out by the parabolic viscous term, and the shear stress reaches its steady state value. This steady state system represents the initial state of the system for a consecutive simulation in which the shearing is stopped and the system is allowed to relax. Fig. 6(b) displays this stress relaxation of the system. The shear stress initially retains its steady state value until velocity at the layer starts to decay. The overshoot observed at $t \approx 7 \times 10^{-1} \tau$ to $t \approx 2 \times 10^{0} \tau$ could be just an artifact of the finite-difference approximation. After this time, the shear stress decays in the fashion of a smeared shock wave, with the amplitude of the shock waves being damped viscously, until it vanishes.

The extent to which an ideal (noninteracting) polymer chain is stretched under a nonuniform flow velocity field can be quantified by calculating the eigenvalues of the tensor of the second moment of the end-to-end vector. The three eigenvalues quantify stretching along three principle axes of this tensor. Fig. 7 displays the time evolution, at a mid-channel layer, of the (a) largest, (b) intermediate and (c) smallest eigenvalues, normalized by their equilibrium values. The chain length is $N_m = 24$ and the walls are neutral ( $\chi_s = 0$ ). The system undergoes a simple shear flow (at a nominal shear rate $\dot{\gamma} = 1 \times 10^{-5} \tau^{-1}$ ). At this shear rate, the stretching of the chain is very small. However, it still can be noticed that, similarly to the time evolution of the flow velocity, both the highest and smallest eigenvalues seem to converge toward their steady state values in a non-monotonic, oscillatory fashion.

Fig. 8(a) displays the time evolution of the first normal stress difference, obtained at $\dot{\gamma} = 1 \times 10^{-5} \tau^{-1}$. The walls are again assumed to be neutral ( $\chi_s = 0$ ). The first normal stress difference is defined by $N_1 = \sigma_{xx} - \sigma_{zz}$ and its positive value indicates that there is a higher degree of orientation in the flow direction (the $x$-direction) than in the direction of the velocity gradient (the $z$-direction). The four different curves in Fig. 8(a) correspond to four different chain lengths $N^A$ = 8, 16, 24 and 32, showing the dependence of $N_1$ on the chain length. The time-dependent growth of $N_1$ upon start-up of steady shearing also follows a non-monotonic, oscillatory trend, until $N_1$ converges to its steady state value. Fig. 8(b) displays an enlarged view of the rise and decay of these oscillations.

The cross-channel variation of the steady-state profile of the component $\langle Q_x Q_z \rangle$ of the tensor of the second moments of the end-to-end vector of an *ideal* (noninteracting) chain, is shown in Fig. 9. The steady-state profile is obtained at a nominal shear rate $\dot{\gamma} = 1 \times 10^{-5} \tau^{-1}$. The component $\langle Q_x Q_z \rangle$ exhibits a higher value at the layers adjacent to the walls, relative to the bulk value. In the next few layers, it approaches the bulk value in a non-monotonic, oscillatory manner. We stress that Fig. 9 represents the direct effect of flow on chain conformations, though if the walls are not neutral, interactions with the walls will influence $\langle Q_x Q_z \rangle$ indirectly. This is because they affect the depletion of segmental volume fractions in the vicinity of the walls (as shown in Fig. 2) and segmental depletion affects the velocity slip in the vicinity of the walls (Figs. 3 and 4). This gives rise to large velocity gradients in the FENE-P dumbbell equation used to evolve $\langle Q_x Q_z \rangle$ in the vicinity of the walls (Eq. (5)).

## IV. PHASE-SEPARATED SYMMETRIC HOMOPOLYMER BLENDS

### A. Simulation Set Up and Parameter Estimation

As the next application of our DSCF theory, we chose to simulate phase-separated, two-component symmetric homopolymer blends in a sheared planar channel. On the one hand, this is a more challenging system to model, since it requires resolution of both transient and steady-state interfacial dynamics, structure and rheology on the scale of the Kuhn length. On the other hand, it gives rise to a characteristic velocity slip and to a reduction of interfacial shear viscosity relative to its value in the bulk phases. This phenomenon has been first predicted based on scaling arguments by de Gennes and co-workers [7,8]. More recently, Goveas and Fredrickson [9] analyzed it using numerical solutions of approximate constitutive equations for the evolution of composition and the deviatoric stress. These were obtained from the Fokker-Plank equation for the Rouse model, assuming fluid incompressibility, guessing an approximate form for the noise and stretching terms in the stress evolution, and neglecting the higher-order terms. Subsequently, Barsky and Robbins published a very detailed molecular dynamics study of this phenomenon [10] and related it to the chain conformation at the sheared interface [11]. They reported a qualitative agreement with the trends predicted by the theoretical studies above, but a significant numerical discrepancy with the results of Goveas and Fredrickson [9]. Most recently, similar velocity slip phenomena have been observed experimentally [12-14], though the bulk phases of the observed polymer fluids were in the entangled regime. Hence the steady-state results of the DSCF study of this phenomenon, which we report in this section, can be compared with previously published results of other theoretical and experimental studies of this phenomenon.

In our DSCF model for sheared symmetric blends, we used polymer chains consisting of identical numbers of freely jointed Kuhn segments below the entanglement length, $N^A = N^B < N_e$. Similarly to Section III, the Kuhn segment length of both components were assumed to be $a = 4.6 \overset{\circ}{\text{A}}$. The glass transition temperatures and the densities of the pure one-component A and B melts at $T = 509 \, ^\circ \text{K}$ and atmospheric pressure were assumed to be identical for both components and their values were chosen to be the same as in Section III. The segmental friction coefficients of the two pure one-component A and B melts were assumed to be identical as well, and their value was chosen to be the same as in Section III. These ideally symmetric blends were first equilibrated in a channel between two stationary walls, kept at a constant temperature $T = 509 \, ^\circ \text{K}$. This was done by integrating the DSCF time-evolution equations using zero wall velocities. The equilibration was initialized using a sharp step function profile of the segmental volume fraction of the two species across the channel, and resulted in a coexistence of a majority-A and a majority-B bulk phases, with a smooth interfacial profiles of the segmental volume fractions of the two species across the interface between the two phases as shown by the solid lines in Fig. 10. Using equilibrium results as an initial input at the time $t = 0$, the DSCF equations are then integrated for $t > 0$ with the top and bottom wall velocities set to constant but opposite values $\mathbf{u}_w = \pm u_w \mathbf{e}_1$, as shown in Fig.

1(b), thus setting the nominal shear rate to $\dot{\gamma} = 2u_w / \left[ \sqrt{2/3} \left( L+1 \right) a \right]$.

## B. Simulation Results for Immiscible Blends

Fig. 10 shows the segmental volume fraction profiles of the phase-separated two-component homopolymer blend, across the channel. The system exhibits coexistence of two bulk phases, A and B, and an interfacial region of finite width between them. Phase A has a majority of A-type homopolymers and a minority of B-type homopolymers. Phase B consists of a majority of B-type chains and a minority of A-type chains. The solid lines denote the equilibrium profiles (corresponding to time $t = 0$). Upon equilibration, the system is sheared at a nominal shear rate $\dot{\gamma} = 1 \times 10^{-3} \tau^{-1}$ until the steady state is achieved. The steady state segmental volume fraction profiles are represented by the symbols. At this shear rate, the steady state profiles are not noticeably different from the equilibrium profiles. The short-range interactions between the polymer segments and the solid walls are neutral ($\chi_s = 0$). For such interactions, the depletion of polymer segments in the vicinity of the wall is quite weak on the scale shown.

As any mean-field theory, our DSCF model neglects thermal fluctuations. Thus the interfacial profile reported in Fig. 10 is an "intrinsic" interfacial profile that neglects capillary wave fluctuations. These are expected to broaden significantly the effective thickness of the interface between the two phases. Hence, for a *quantitative* comparison of interfacial with experimental observations and with the results of MD and Monte Carlo simulations, it is necessary to account for the broadening of the intrinsic profile obtained from DSCF by capillary wave fluctuations, using theoretical estimates of their spectrum [34-37]. We report such a *quantitative* comparison of DSCF and MD interfacial profiles, accounting for capillary wave broadening, in a separate work [38].

Fig. 11 displays the cross-channel variation of $u_x$, the $x$-component of the mass-averaged steady state velocity, centered about the polymer-polymer interface ($z = 0$), obtained at a nominal shear rate $\dot{\gamma} = 1 \times 10^{-3} \tau^{-1}$. There is a significant upward kink in the velocity profile in the interfacial region. This phenomenon, called interfacial velocity slip, has been the subject of a number of theoretical [7-9], MD [10] and experimental [12-14] studies, as surveyed above. It can be quantified by the intercepts of the extrapolation of linear portion of the velocity profiles with the velocity axis (determining a velocity slip), and with the $z$-axis normal to the interface (determining a slip length). The velocity slip in Fig. 11 is qualitatively very similar to the velocity slip reported in the references above.

The time evolution of the shear stress, obtained at a nominal shear rate $\dot{\gamma} = 1 \times 10^{-5} \tau^{-1}$, is shown in Fig. 12. We consider here the development of the shear stress, after the onset of steady shear, at three distinct layers: the one next to the lower wall (solid line), the one at the lower quarter-channel corresponding to a bulk phase (dotted line), and the layer closest to the middle of the interfacial region (dashed line). Since the momentum propagates from the walls, the shear stress rises faster at the layer next to the wall, relative to the other two layers. At small times ($\approx 10^0 \tau$), it seems that the shear stress exhibits a slight undershoot, before rising. This artifact seems unphysical and is perhaps related our use of the finite-difference approximations [29]. All three profiles indicate that the shear stress develops as a shock wave, with the amplitude decaying due to viscous loss. The shock wave is being reflected back and forth by the walls and the interfacial region. It is well known that simple finite-difference approximation schemes tend to smooth out shock waves [29], which thus resemble oscillations in Fig. 12. Note that the amplitude of the oscillations is much smaller at the layer in a bulk phase than at the other two layers. After a while, these oscillations disappear completely and the shear stress achieves its steady state value. Such development of the shear stress is related to the non-monotonic velocity propagation of dampened oscillations, or smeared shock-waves, as was seen in our DSCF simulations of one-component melts as described in Section III.

The local stretching of an ideal (noninteracting) polymer chain by nonuniform flows can be quantified by calculating the eigenvalues of the tensor of the second moment of the end-to-end vector for ideal chains, with their center of mass located at a particular layer. The three eigenvalues measure stretching along the three principle axes of this tensor. Fig. 13 shows the variation across the channel of the (a) largest, (b) intermediate, and (c) smallest eigenvalues at the steady state, normalized by their equilibrium values. The steady state is obtained at a nominal shear rate $\dot{\gamma} = 1 \times 10^{-3} \tau^{-1}$. At this shear rate, there is a significant stretching of the chain, especially near the walls and at the polymer-polymer interface. However, here we consider an ideal (noninteracting) chain and this stretching is only due to the nonuniform steady state velocity profile. It does not account for segment-segment and segment-wall interactions. These are modeled by the interactions of segments with a self-consistent field as they perform random walks generating a chain configuration. Though the effects of these interactions are included in our DSCF evolution equations, we are not aware of a method to calculate the second moment of the end-to-end distance of *interacting* chains *directly* from the DSCF evolution equations. However, the free-segment and stepping probabilities obtained by numerical solutions of the DSCF evolution equations can be used to provide transition rates for *Monte Carlo sampling* of the second moment of the end-to-end distance of *interacting* chains. Such Monte-Carlo DSCF calculations have been presented in a separate work [38].

Fig. 14 compares the time evolution of the first normal stress difference $N_1$ in a bulk phase (quarter-channel layer) and at the interface (mid-channel layer), obtained at $\dot{\gamma} = 1 \times 10^{-5} \tau^{-1}$. These two profiles indicate that the local first normal stress difference $N_1$ grows in a non-monotonic, dampened oscillatory fashion, until it converges to the steady state value. Again, these dampened oscillations could also be interpreted as dampened shock-waves being reflected back and forth between the two walls and the interface, smoothed over by the finite-difference approximation scheme to the governing partial differential equation, which is of hyperbolic nature [29]. The amplitude of oscillations is higher, and decays slower at the interface than in the bulk. At the steady state, $N_1$ is larger in the bulk than at the interface.

Fig. 15 shows the variation of the local shear viscosity across the interface, calculated by dividing the local shear stress, obtained from the local momentum density flux, by the local shear rate obtained, from the gradient of the velocity. In this figure, the interface is centered about the plane $z = 0$, and the local shear viscosity decreases from the bulk values and reaches a minimum as $z = 0$ is approached. Similar decrease in interfacial viscosities was reported in studies using the scaling approach [7,8], numerical solutions of approximate constitutive equations obtained from the Fokker-Plank equations for the Rouse model [9], molecular dynamics simulations of bead-spring models [10], and experimental studies [12-14].

## V. CONCLUSIONS

We reported here the first application of our novel DSCF lattice theory for unentangled, inhomogeneous polymer fluids in a planar, steadily sheared channel. Starting from the general three-dimensional DSCF theory presented in the preceding paper [15], simplified quasi-one-dimensional DSCF equations were derived for systems exhibiting translational invariance within planes parallel to the sheared walls. The quasi-1D DSCF equations were then applied to one-component homopolymer melts, and to phase-separated, symmetric, two-component homopolymer blends in a planar, steadily sheared channel possessing such translational symmetry. The quasi-1D DSCF equations were used to calculate the time evolution and the steady-state profiles of composition, density, velocity, chain deformation, stress, viscosity and normal stress within layers across the sheared channel in sheared melts and blends. Good qualitative agreement is obtained with previously observed phenomena, such as chain depletion next to the walls and interfaces, and the associated velocity slip and reduced viscosity in interfacial regions of symmetric immiscible blends.

A detailed *quantitative* comparison of our DSCF approach with molecular dynamics simulations of unentangled inhomogeneous polymer fluids is beyond the scope of this work. Such quantitative analysis requires conversions between the system of units used in the DSCF model and the system of units used in the model used in the MD simulations. It also has to account for the effective broadening of the intrinsic interfacial profiles obtained from the mean-field DSCF equations, caused by capillary wave fluctuations [34-37]. Finally, quantitative analysis of the second-moment of the end-to-end distance of *interacting* chains require Monte Carlo simulations of conformations of single chains as random walks in a self-consistent field, where the transition rates for this Markov process are provided by the DSCF evolution equations. We have recently succeeded to make such a quantitative comparison in a recent collaboration with Wentao Li and Dilip Gersappe (SUNY at Stony Brook), who produced the molecular Dynamics simulations. The results, which will be reported in a separate publication [38], validate our DSCF method for unentangled, inhomogeneous polymer fluids.

The first applications of the DSCF approach presented here indicate that it may be possible to use it to model interfacial dynamics of a rich variety of inhomogeneous fluid that were extensively studied at equilibrium using static SCF methods. We intend to pursue this goal in subsequent publications.

## ACKNOWLEDGEMENTS

We acknowledge numerous helpful discussions with Drs. Glen H. Ko, Stanislav Solovyov, Wen Tao Li and Dilip Gersappe, for which we are very grateful. We thank Dr. Toshihiro Kawakatsu for sending us a preprint [39] summarizing the relevant work of his group prior to its publication. Most of this work was performed while Y. Shnidman was a member of the Department of Chemical Engineering and Chemistry at the Polytechnic University in Brooklyn, NY, where Maja Mihajlovic and Tak Shing Lo were a Ph. D. candidate and a postdoctoral research associate in his group, respectively. Some preliminary results were included in the Ph. D. thesis [40] submitted by Maja Mihajlovic to Polytechnic University in Brooklyn, NY, in partial fulfillment of requirements for the Ph. D. degree in Chemical Engineering. Maja Mihajlovic and Yitzhak Shnidman thank the National Science Foundation for financial support (grant DMR-0080604). Tak Shing Lo and Yitzhak Shnidman acknowledge a grant from the Mitsubishi Chemical Corporation of Japan, for which they are grateful.

**Fig. 1**

Inhomogeneous polymer fluids in a channel between two solid walls sheared along the x-axis at opposite velocities $\mathbf{u}_w = \pm u_w \left(1,0,0\right)$, with a nominal shear rate $\dot{\gamma}$. Physical properties are expected to vary along the z-axis, but stay invariant in layers parallel to the walls (dashed lines). Chain conformations are shown as biased random walks (solid white lines) in a self-consistent field. (a) A melt of a single species. (b) A phase-separated blend of two species A and B, with the interface between the majority-A and majority B species that is parallel to the wall and centered at mid-channel. (c). Enlarged view of a site (labeled 0) in a triangular lattice layer $i$ on the FCC lattice, surrounded by all its nearest neighbors, labeled 1 to 6 in the same layer, 7 to 9 in the triangular lattice layer $i+1$, and 10 to 12 in the triangular lattice layer $i-1$.

**Fig. 2**

Steady state segmental volume fraction profiles of a one-component homopolymer melt, across the channel, obtained at a nominal shear rate $\dot{\gamma} = 1 \times 10^{-3} \tau^{-1}$. The polymer chain consists of $N^A = 24$ Kuhn segments. The short-range interactions between the polymer segments and the solid walls vary between attractive ($\chi_s > 0$), neutral ($\chi_s = 0$) and repulsive ($\chi_s < 0$).

**Fig. 3**

Time evolution of the velocity profiles across the channel. The initial state of the system (at time $t = 0$) corresponds to the system at equilibrium. At subsequent times, the system is sheared at a nominal shear rate $\dot{\gamma} = 1 \times 10^{-5}\ \tau^{-1}$. The walls are neutral ($\chi_s = 0$), and the chain length is $N^A = 24$.

**Fig. 4**

Time evolution of the velocity in the lower quarter-channel, obtained at a nominal shear rate $\dot{\gamma} = 1 \times 10^{-5}\ \tau^{-1}$. The walls are neutral ($\chi_s = 0$), and the chain length is $N^A = 24$.

**Fig. 5**

Velocity profiles across the channel, given at selected times and at steady state (denoted by the dash-dotted line). The system is sheared at a nominal shear rate $\dot{\gamma} = 1 \times 10^{-5}\ \tau^{-1}$. The walls are neutral ($\chi_s = 0$), and the chain length is $N^A = 24$.

**Fig. 6**

Time evolution of the shear stress, at the lower quarter-channel, following: (a) the onset of a steady shear, and (b) the secession of the shear. The nominal shear rate is $\dot{\gamma} = 1 \times 10^{-5}\ \tau^{-1}$. The walls are neutral ($\chi_s = 0$), and the chain length is $N^A = 24$.

**Fig. 7**

Time evolution of the stretching of an ideal (non-interacting) polymer chain under flow (at $\dot{\gamma} = 1 \times 10^{-5}\ \tau^{-1}$). The stretching is calculated using the (a) largest, (b) intermediate and (c) smallest eigenvalues of the second moment of the end-to-end distance, normalized by their values at equilibrium. The walls are neutral ($\chi_s = 0$), and the chain length is $N^A = 24$.

**Fig. 8**

(a) Time evolution of the first normal stress difference, $N_1 = \sigma_{xx} - \sigma_{zz}$, obtained at a nominal shear rate $\dot{\gamma} = 1 \times 10^{-5} \ \tau^{-1}$. The chain length varies from $N^A = 8$ to $N^A = 24$. (b) An enlarged view of the rise of the first normal stress difference. The walls are neutral ($\chi_s = 0$).

**Fig. 9**

The $\langle Q_x Q_z \rangle$ component of the second moment of the end-to-end vector of the FENE-P dumbbell across the channel, at steady state. The steady state is achieved at a nominal shear rate $\dot{\gamma} = 1 \times 10^{-5} \ \tau^{-1}$. The walls are neutral ($\chi_s = 0$), and the chain length is $N^A = 24$.

**Fig. 10**

Steady state segmental volume fraction profiles of a two-component polymer blend, across the channel, obtained at a nominal shear rate $\dot{\gamma} = 1 \cdot 10^{-3} \ \tau^{-1}$. Polymer chains consist of $N^A = N^B = 24$ Kuhn segments. The Flory-Huggins interaction parameter is $\chi_{AB} = 0.55$ and the walls are neutral ($\chi_s = 0$).

**Fig. 11**

Steady state velocity profile, centered about the planar interface, of a phase separated, two-component polymer blend, obtained at $\dot{\gamma} = 1 \cdot 10^{-3} \ \tau^{-1}$. Polymer chains consist of $N^A = N^B = 24$ Kuhn segments. The Flory-Huggins interaction parameter is $\chi_{AB} = 0.55$, and the walls are neutral ($\chi_s = 0$).

**Fig. 12**

Time evolution of shear stress, at selected layers, obtained at $\dot{\gamma} = 1 \cdot 10^{-5}$ $\tau^{-1}$. Polymer chains consist of $N^A = N^B = 24$ Kuhn segments. The Flory-Huggins interaction parameter is $\chi_{AB} = 0.55$ and the walls are neutral ($\chi_s = 0$).

**Fig. 13**

Stretching of an ideal (noninteracting) polymer chain at steady state (obtained at $\dot{\gamma} = 10^{-3}$ $\tau^{-1}$), across the channel. The stretching is calculated using the (a) largest, (b) intermediate and (c) smallest eigenvalues of the second moment of the end-to-end distance, normalized by their values at equilibrium. The chain length is $N^A = N^B = 24$. The Flory-Huggins interaction parameter is $\chi_{AB} = 0.55$ and the walls are neutral ($\chi_s = 0$).

**Fig. 14**

Time evolution of first normal stress difference, in a bulk phase and at the interface. A nominal shear rate is $\dot{\gamma} = 10^{-5}$ $\tau^{-1}$. The chain length is $N^A = N^B = 24$. The Flory-Huggins interaction parameter is $\chi_{AB} = 0.55$ and the walls are neutral ($\chi_s = 0$).

**Fig. 15**

Steady state viscosity profile, centered about the interface, of a phase separated, two-component polymer blend, obtained at $\dot{\gamma} = 1 \cdot 10^{-3}$ $\tau^{-1}$. Polymer chains consist of $N^A = N^B = 24$ Kuhn segments. The Flory-Huggins interaction parameter is $\chi_{AB} = 0.55$ and the walls are neutral ($\chi_s = 0$).

Fig. 1

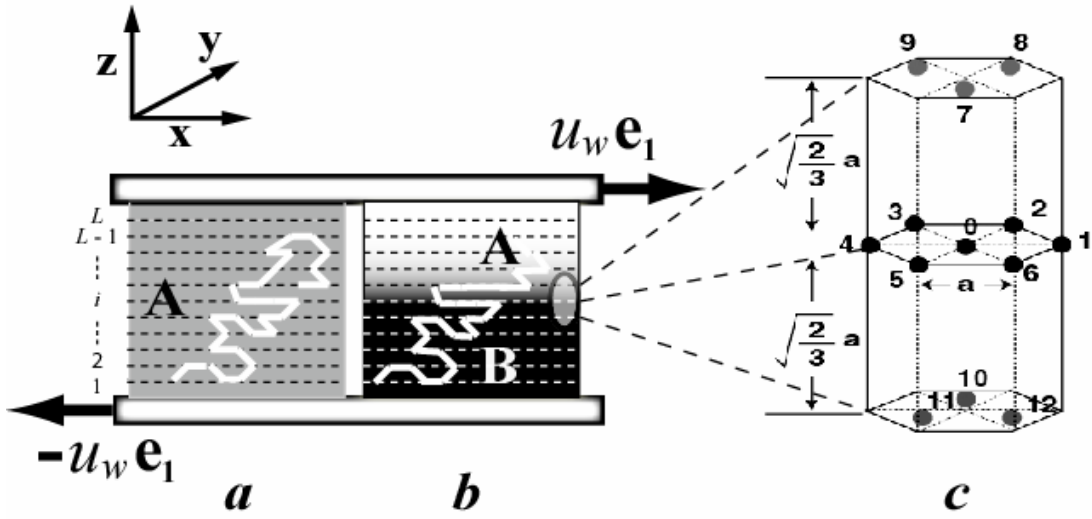

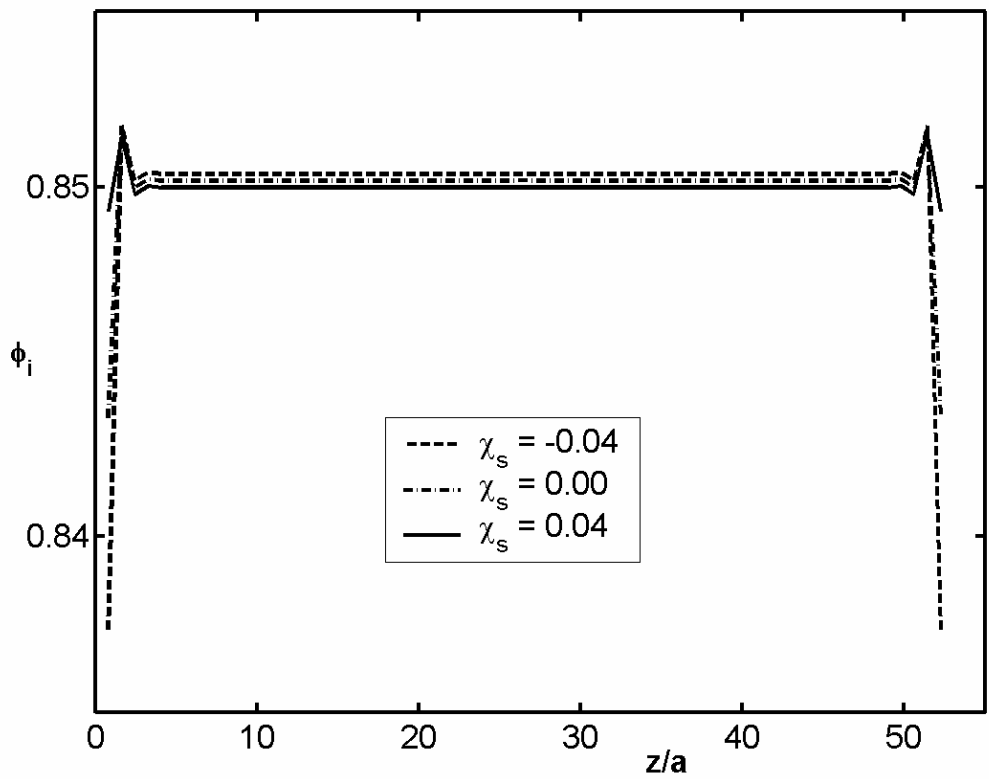

Fig. 3

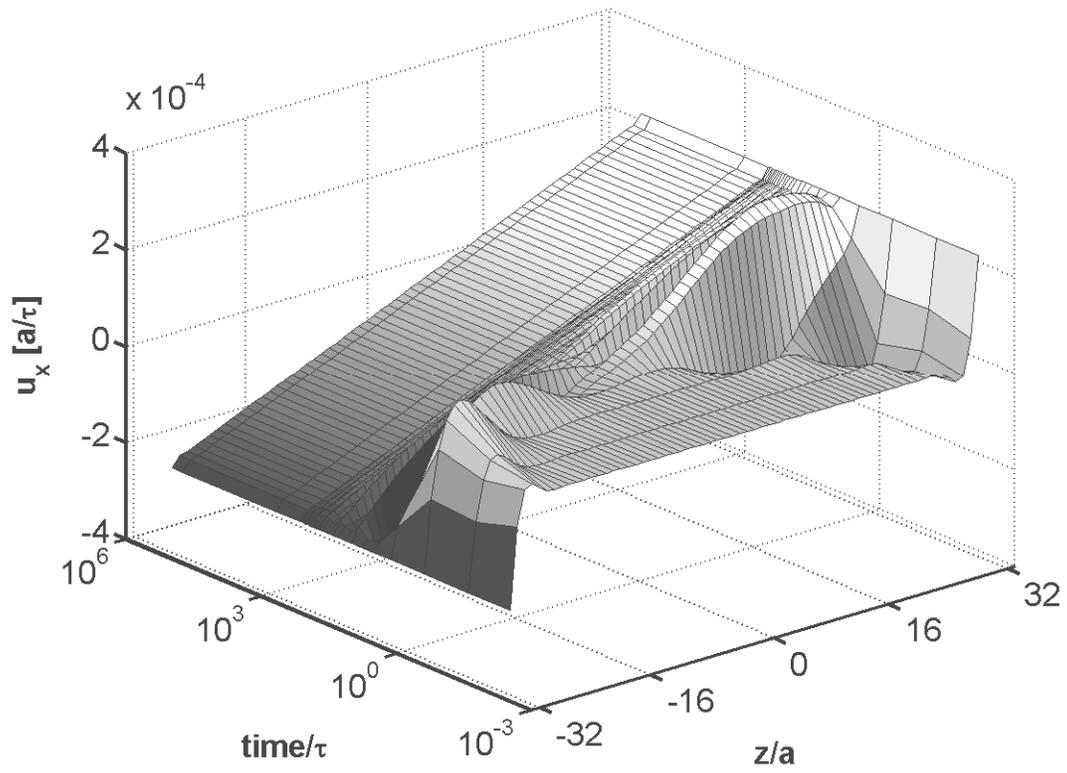

Fig. 4

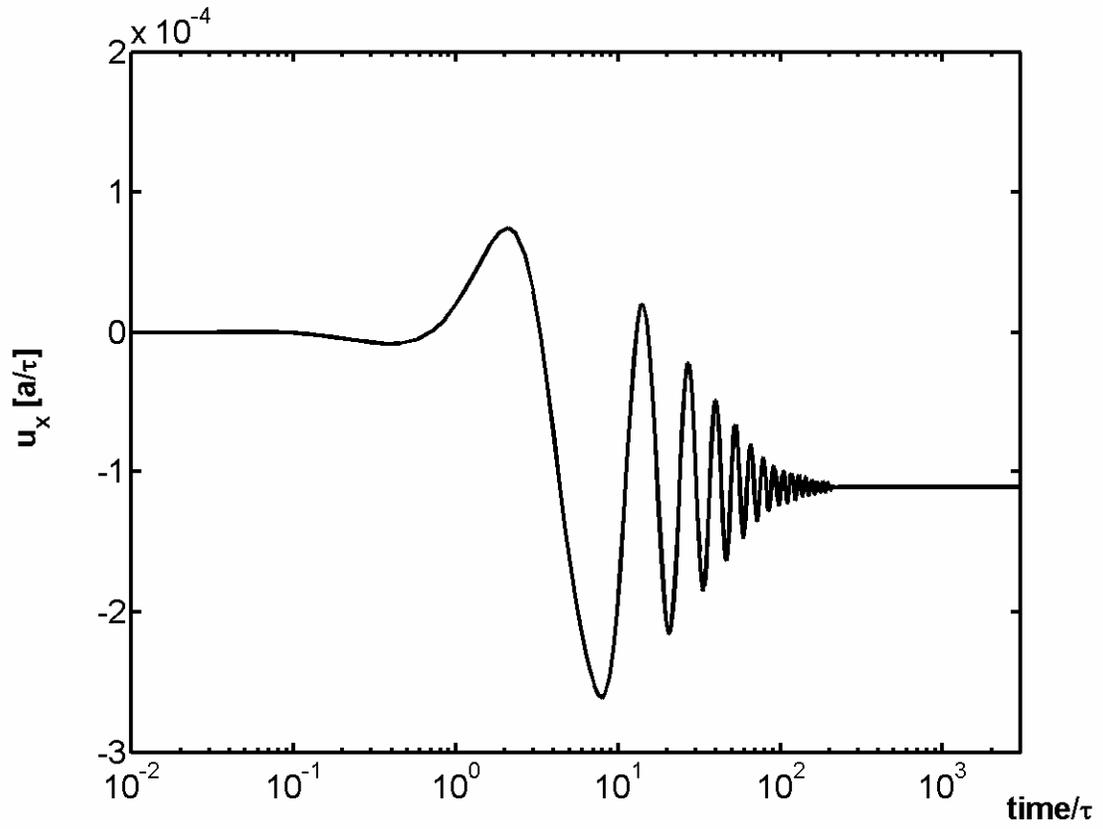

Fig. 5

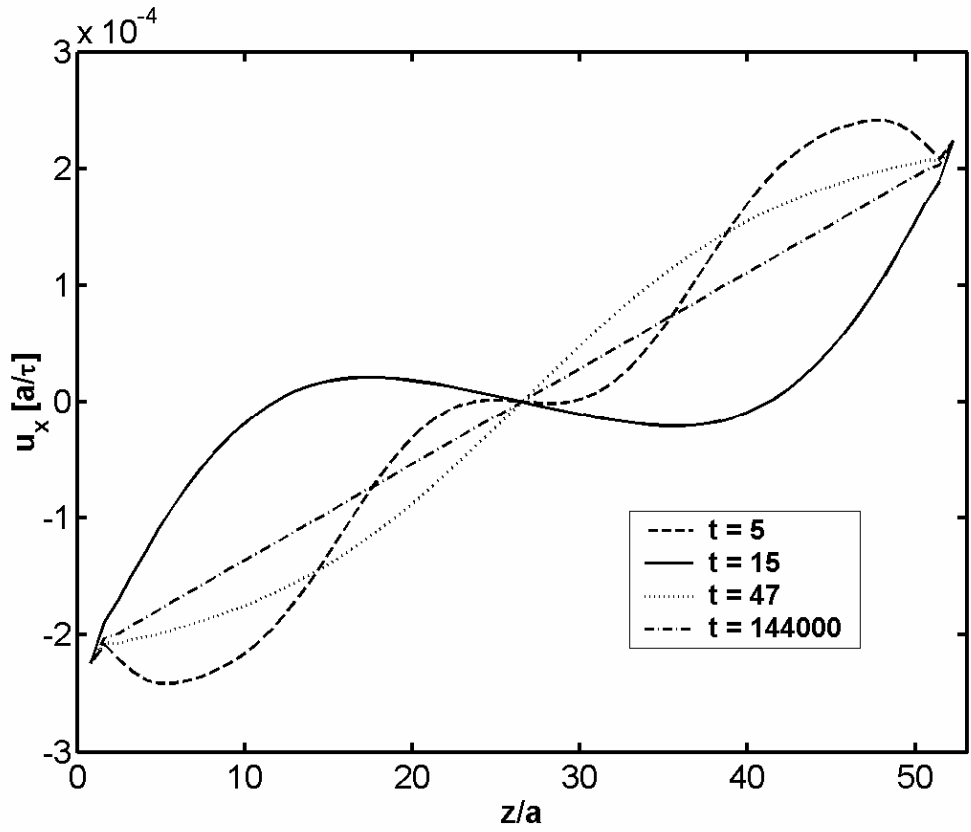

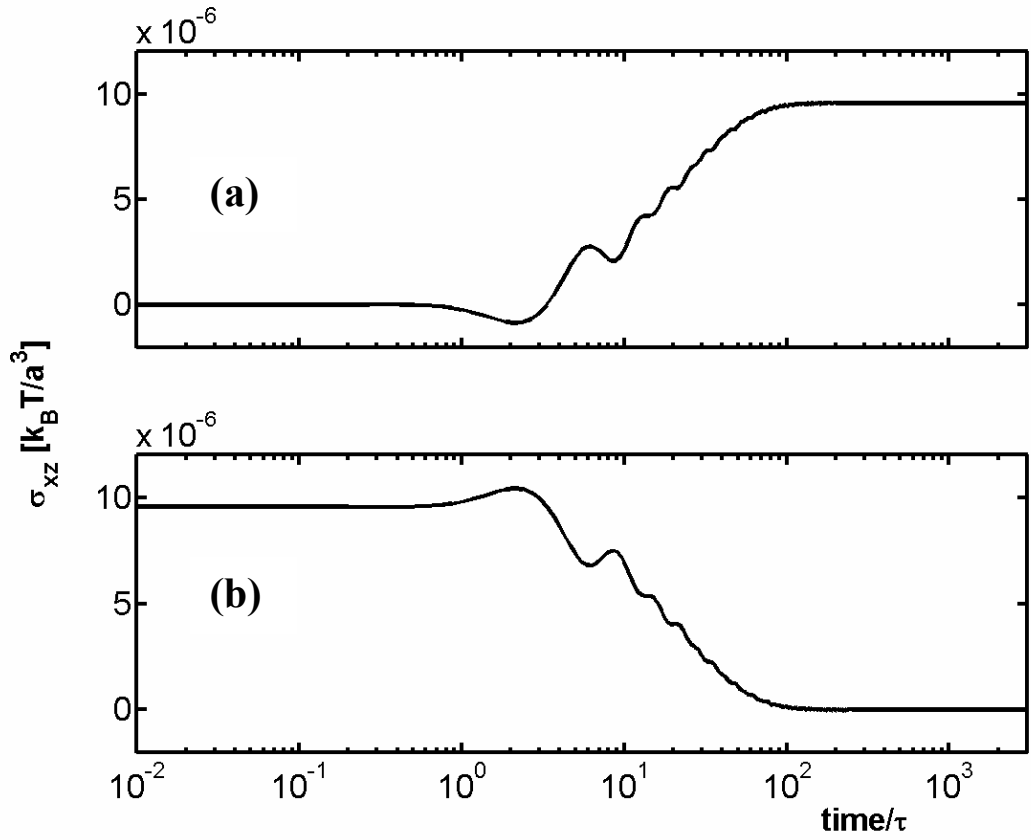

Fig. 7

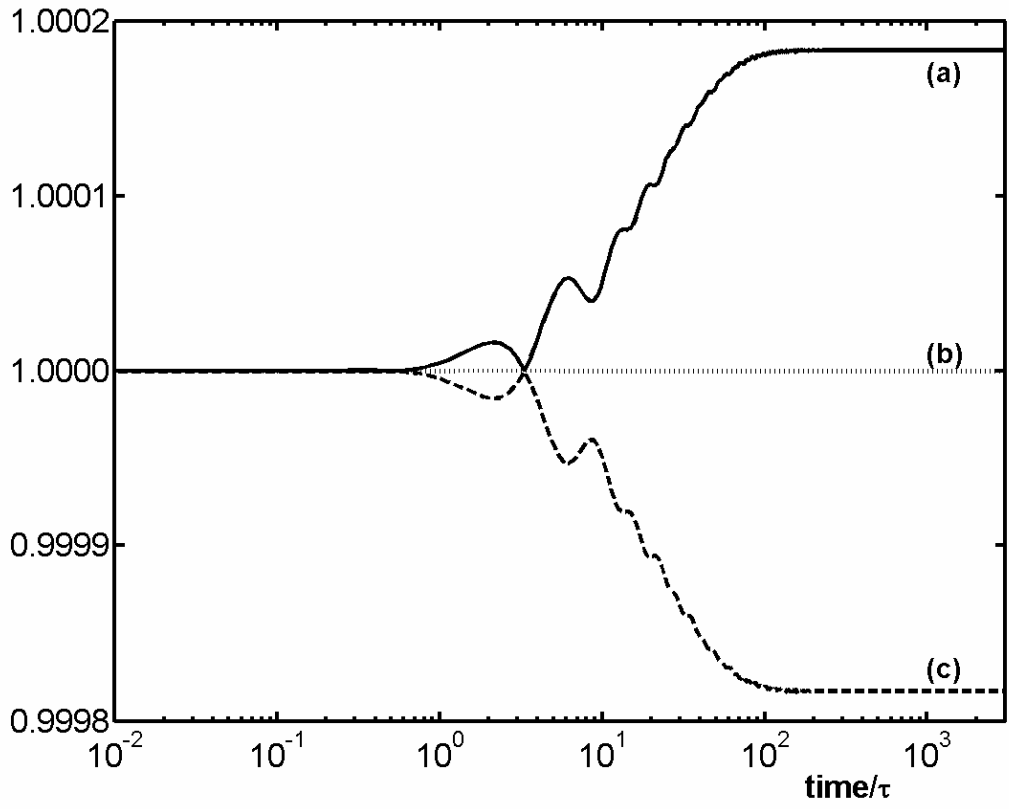

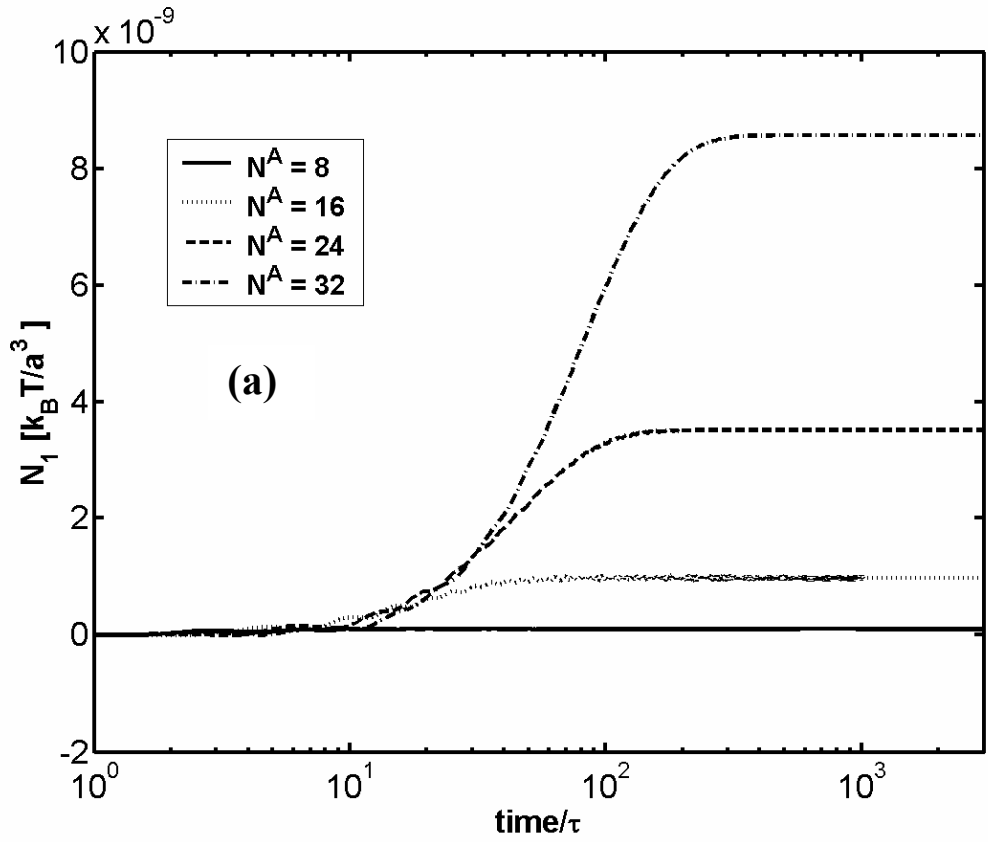

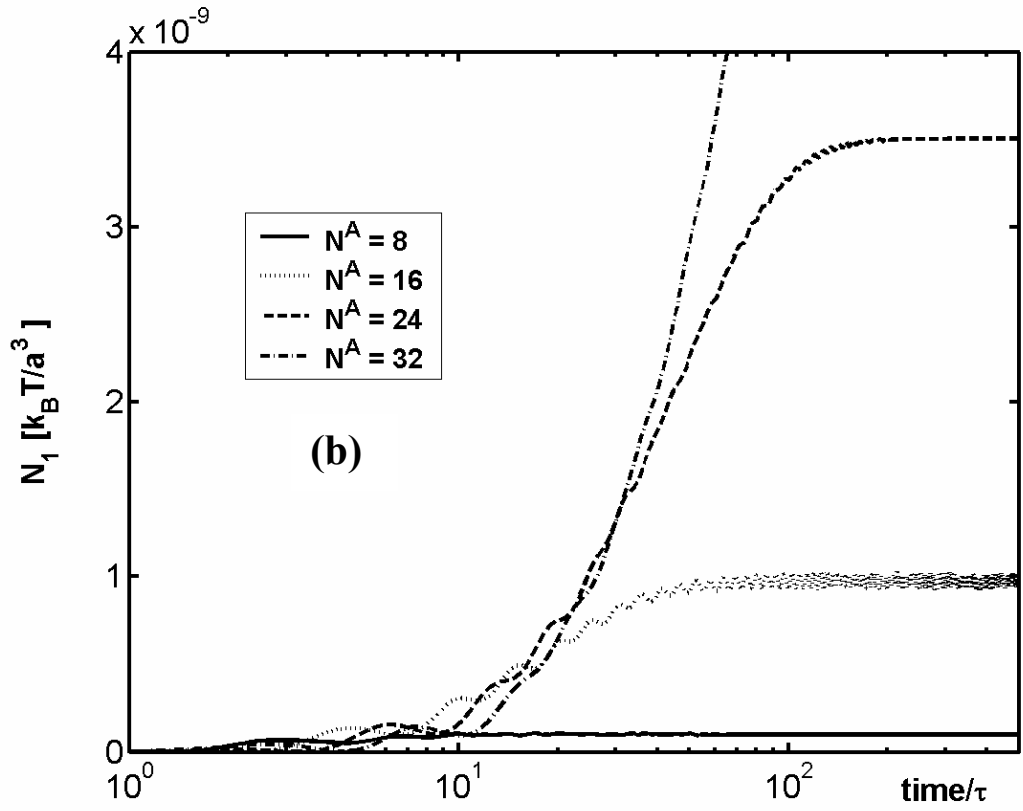

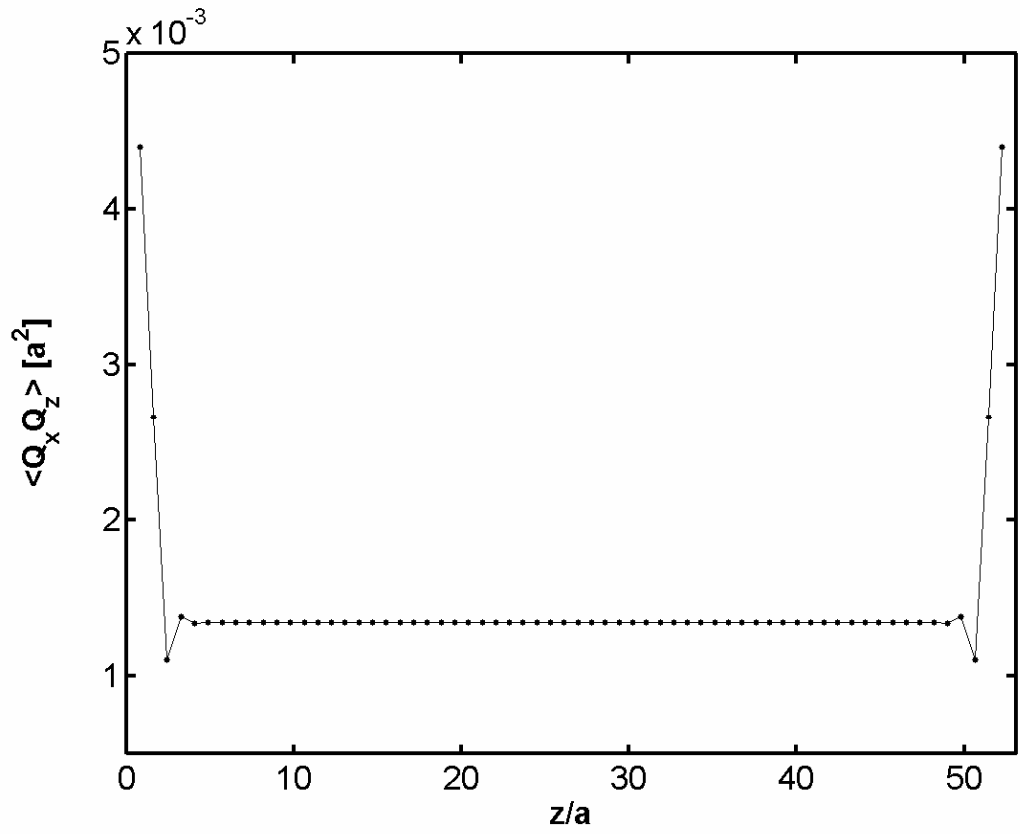

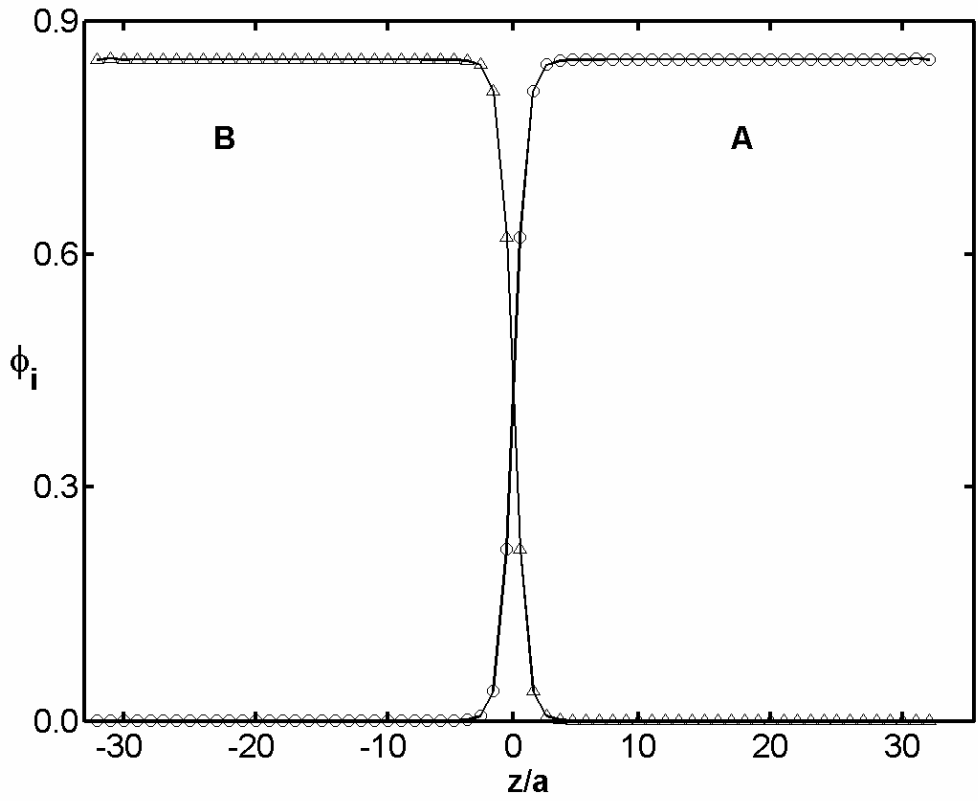

Fig. 11

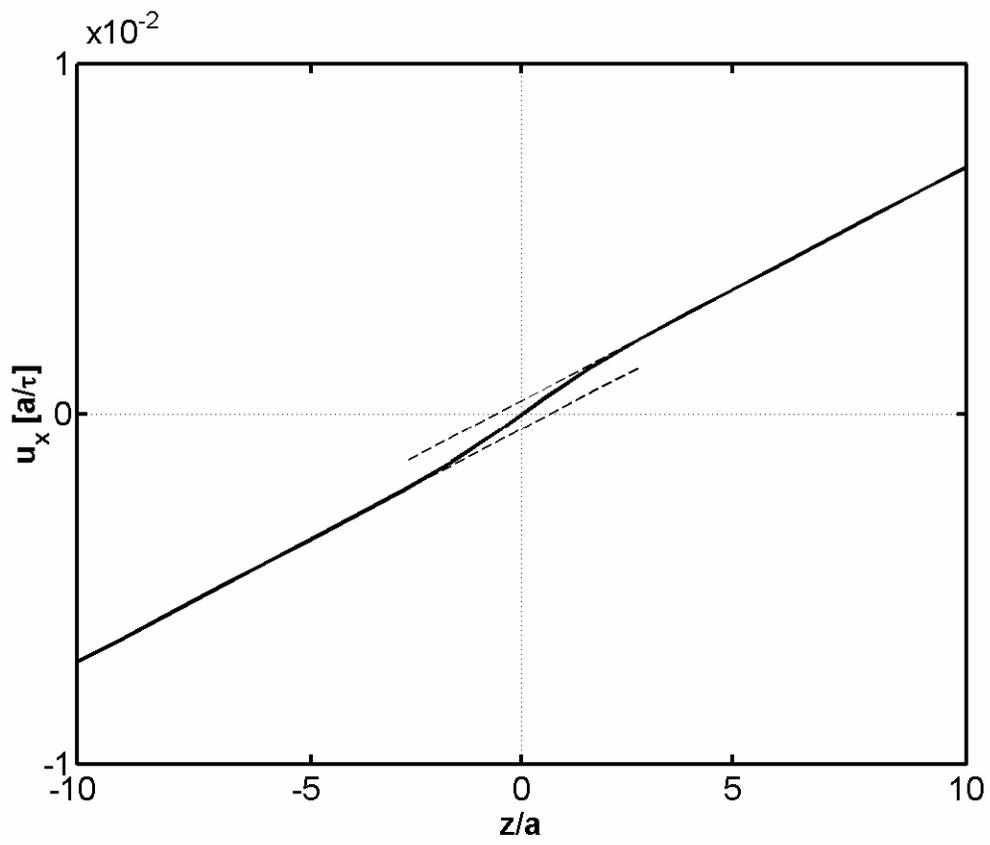

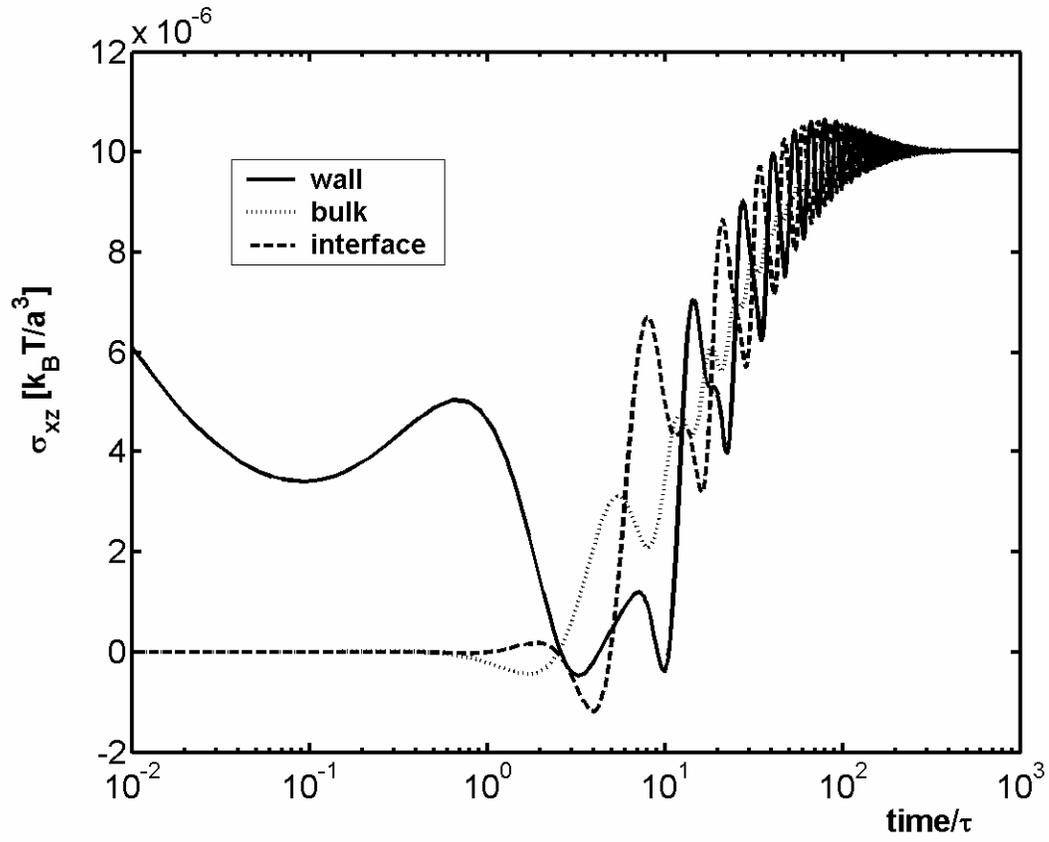

Fig. 13

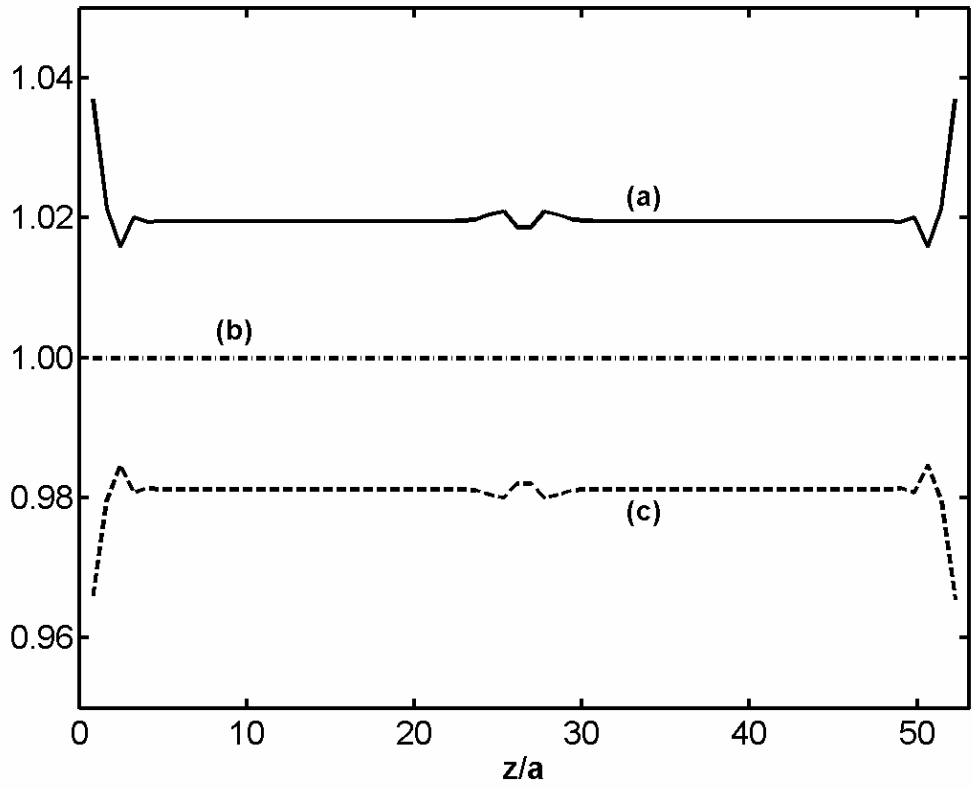

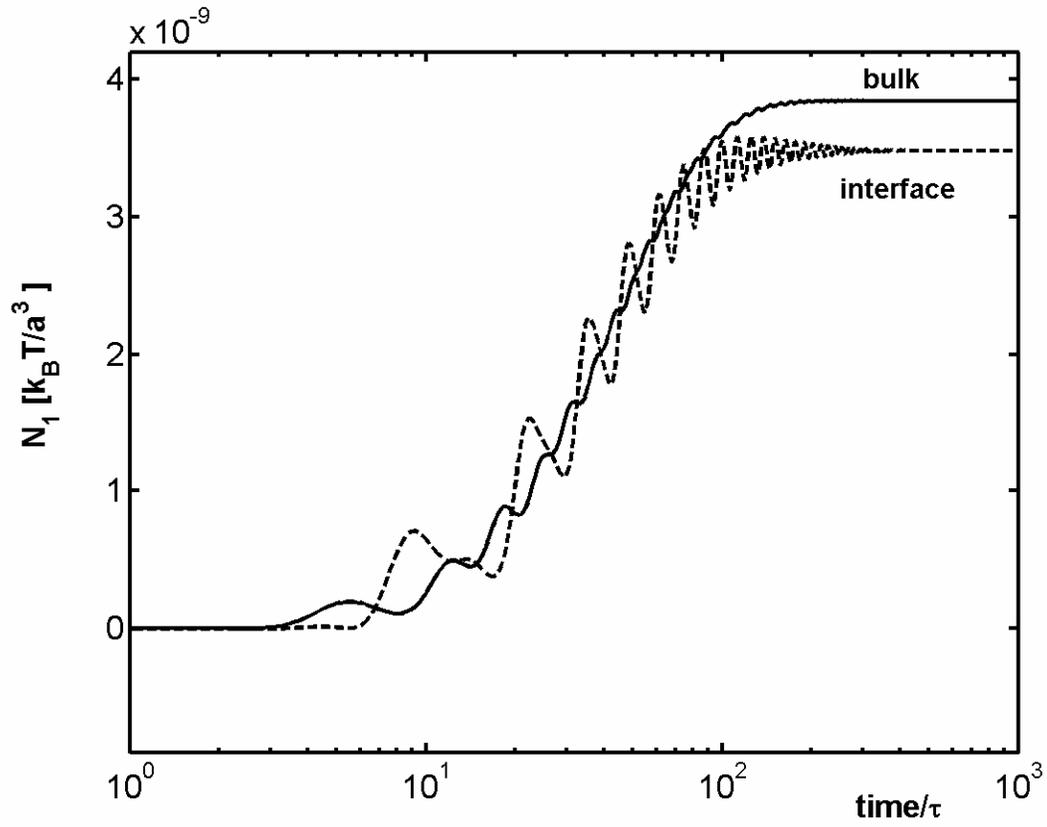

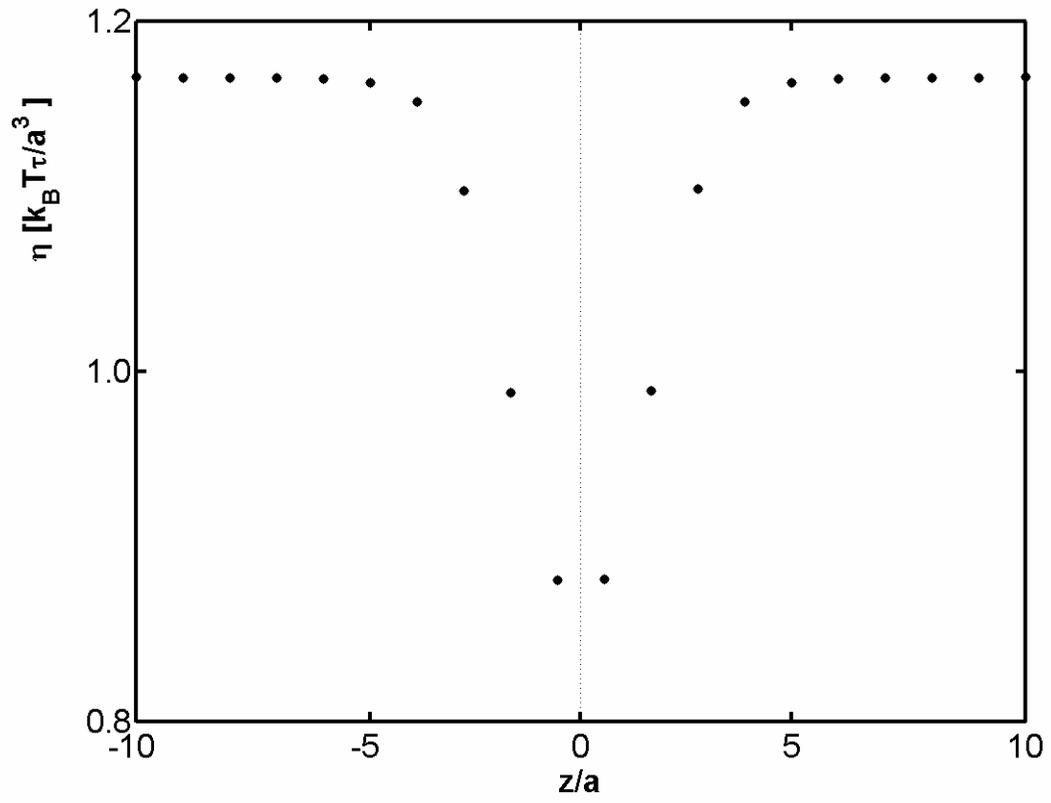

# References

[1] M. Doi and S. F. Edwards, *The Theory of Polymer Dynamics* (Oxford University Press, Oxford, 1986).

[2] R. B. Bird, C. F. Curtiss, R. C. Armstrong, and O. Hassager, *Dynamics of Polymeric Fluids*, 1987.

[3] R. G. Larson, *Constitutive Equations for Polymer Melts and Solutions* (Butterworths, Boston, 1988).

[4] R. G. Larson, *The Structure and Rheology of Complex Fluids* (Oxford University Press, Oxford, 1999).

[5] T. C. B. McLeish, Adv. Phys. **51**, 1379 (2002).

[6] M. M. Denn, Annu. Rev. Fluid Mech. **33**, 265 (2001).

[7] P. G. de Gennes, C. R. Acad. Sci. Paris, Ser. II **308**, 1401 (1989).

[8] F. Brochard-Wyart, P. G. de Gennes, and S. Troian, C. R. Acad. Sci. Paris, Ser. II **310**, 1169 (1990).

[9] J. L. Goveas and G. H. Fredrickson, Eur. Phys. J. B **2**, 79 (1998).

[10] S. Barsky and M. O. Robbins, Phys. Rev. E **63**, 21801 (2001).

[11] S. Barsky and M. O. Robbins, Phys. Rev. E **65**, 021808 (2002).

[12] R. Zhao and C. W. Macosko, J. Rheol. **46**, 145 (2002).

[13] Y. C. Lam, L. Jiang, C. Y. Yue, K. C. Tam, L. Li, and X. Hu, J. Rheol. **47**, 795 (2003).

[14] Y. C. Lam, C. Y. Yue, Y. X. Yang, K. C. Tam, and X. Hu, J. Appl. Polym. Sci. **87**, 258 (2003).

[15] M. Mihajlovic, T. S. Lo, and Y. Shnidman, Phys. Rev. E, submitted, preceding paper.

[16] J. Scheutjens and G. J. Fleer, J. Phys. Chem. **83**, 1619 (1979).

[17] J. M. H. M. Scheutjens and G. J. Fleer, J. Phys. Chem. **83**, 1619 (1979).

[18] D. N. Theodorou, J. Chem. Phys. **22**, 4578 (1989).

[19] D. N. Theodorou, Macromolecules **22**, 4589 (1989).

[20] A. Hariharan, S. K. Kumar, and T. P. Russell, J. Chem. Phys. **99**, 4041 (1993).

[21] A. K. Doolittle and D. B. Doolittle, J. Appl. Phys. **28**, 901 (1957).

[22] D. Jou, J. Casas-Vasquez, and M. Criado-Sancho, *Thermodynamics of Fluids Under Flow* (Springer, Berlin, 2001).

[23] B. Q. Li and E. Ruckenstein, J. of Chem. Phys. **106**, 280 (1997).

[24] M. Putz, K. Kremer, and G. S. Grest, Europhys. Lett. **49**, 735 (2000).

[25] W. Paul, G. D. Smith, D. Y. Yoon, B. Farago, S. Rathgeber, A. Zirkel, L. Willner, and D. Richter, Phys. Rev. Lett. **80**, 2346 (1998).

[26] A. A. Khan and Y. Shnidman, Prog. Coll. Polym Sci. **103**, 251 (1997).

[27] A. A. Khan, *Lattice-Gas Models of Interfacial Dynamics, Thesis* (University of Rochester, Rochester, NY, 1999).

[28] Y. J. A. Mochimaru, J. Non-Newtonian Fluid Mech. **12**, 135 (1983).

[29] R. J. LeVeque, *Numerical methods for conservation laws* (Birkhauser, Basel, 1992).

[30] D. M. Tolstoi, Dokl. Akad. Nauk SSSR **85**, 1089 (1952).

[31] T. D. Blake, Coll. Surf. **47**, 135 (1990).

[32] E. Ruckenstein and N. Churayev, J. Colloid Interface Sci. **147**, 535 (1991).

[33] A. A. Alexander and O. I. Vinogradova, Coll. Surf. **108**, 173 (1996).

[34] A. N. Semenov, Macromolecules **27**, 2732 (1994).

[35] A. Werner, F. Schmid, M. Muller, and K. Binder, J. Chem. Phys. **107**, 8175 (1997).

[36] M. D. Lacasse, G. S. Grest, and A. Levine, J. Phys. Rev. Lett. **80**, 309 (1998).

[37] K. Binder, M. Muller, F. Schmid, and A. Werner, Adv. Coll. Interf. Sci. **94**, 237 (2001).

[38] T. S. Lo, M. Mihajlovic, Y. Shnidman, W. T. Li, and D. Gersappe, Phys. Rev. Lett., submitted.

[39] T. Shima, H. Kuni, Y. Okabe, M. Doi, H. F. Yuan, and T. Kawakatsu, Macromolecules **36**, 9199 (2003).

[40] M. Mihajlovic, Dynamic Self-Consistent Field Theory of Inhomogeneous Complex fluids under Shear, Thesis (Polytechnic University, Brooklyn, 2004).